\documentclass[12pt,a4paper]{article}
\usepackage{amsmath}
\usepackage{graphicx, psfrag, epsf, orcidlink, setspace}
\usepackage{enumerate}
\usepackage[font=small]{caption}
\usepackage[round, authoryear]{natbib}
\usepackage{url, paralist} 
\usepackage[top=1.6in, bottom=1.6in, left=1.1in, right=1.1in]{geometry}
\usepackage{float,bm,booktabs,makecell,amsthm,amssymb,amsfonts,dsfont,mathtools,authblk,subcaption}

\DeclareMathOperator*{\argmin}{arg\,min}

\usepackage{algorithm}
\usepackage{algorithmic}
\allowdisplaybreaks

\usepackage[font=small]{caption}
\graphicspath{{plots/}}

\usepackage{amssymb}
\usepackage{hyperref}
\usepackage{multirow}
\usepackage{booktabs}
\usepackage{mathrsfs, xcolor} 
\definecolor{darkblue}{rgb}{0,0,.6}
\hypersetup{colorlinks = true, linkcolor=darkblue, urlcolor=darkblue, citecolor=darkblue}
\usepackage{xr}
\usepackage[patch=none]{microtype}

\newcommand{\blind}{0}

\date{}
\newcommand{\Rlogo}{\protect\includegraphics[height=1.8ex,keepaspectratio]{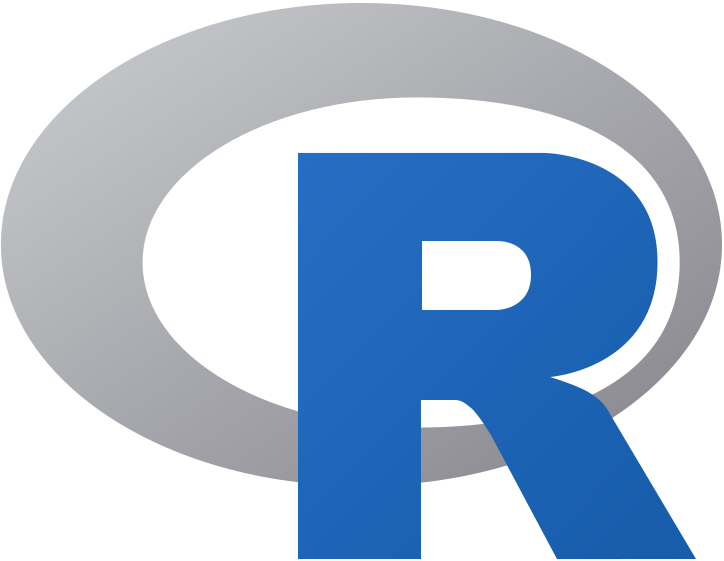}}

\begin{document}

\def\spacingset#1{\renewcommand{\baselinestretch}%
{#1}\small\normalsize} \spacingset{1}

\def\be{\begin{equation}}
\def\ee{\end{equation}} 
\def\ben{\begin{equation*}}
\def\een{\end{equation*}}
\def\bea{\begin{eqnarray}}
\def\eea{\end{eqnarray}}
\def\bda{\begin{eqnarray*}}
\def\eda{\end{eqnarray*}}
\numberwithin{equation}{section}

\newtheorem{definition}{Definition}
\newtheorem{theorem}{Theorem}
\newtheorem{proposition}{Proposition}
\newtheorem{corollary}{Corollary}
\newtheorem{assumption}{Assumption}
\renewcommand\theassumption{A\arabic{assumption}}
\newtheorem{lemma}{Lemma}
\newtheorem*{remark}{Remark}

\theoremstyle{definition}
\newtheorem{exmp}{Example}[section]
\AtEndDocument{\refstepcounter{theorem}\label{finalthm}}
\AtEndDocument{\refstepcounter{proposition}\label{finalprop}}
\newcommand{\pkg}[1]{{\normalfont\fontseries{b}\selectfont #1}} \let\proglang=\textsf \let\code=\texttt
\setlength{\abovedisplayskip}{4pt}
\setlength{\belowdisplayskip}{4pt}

\if0\blind
{
  \title{\bf Spherically Embedded Time Series with Unknown Trend and Periodic Components}
  \author{\normalsize Jiazhen Xu \orcidlink{0009-0006-7870-0340}\footnote{Corresponding author: Department of Actuarial Studies and Business Analytics, Level 7, 4 Eastern Road, Macquarie University, Sydney, NSW 2109, Australia; Email: jiazhen.xu@mq.edu.au} \qquad Han Lin Shang \orcidlink{0000-0003-1769-6430} \\
  \normalsize Department of Actuarial Studies and Business Analytics \\
  \normalsize Macquarie University}
  \date{}
  \maketitle
} \fi

\if1\blind
{
  \title{\bf Spherically Embedded Time Series with Unknown Trend and Periodic Components}
  \author{}
  \maketitle
  \medskip
} \fi

\bigskip

\begin{abstract}
Spherically embedded time series with values naturally residing on or can be equivalently mapped to the sphere often exhibit complex non-stationarity driven by latent trend and periodic components. Traditional Euclidean methods fail to account for the sphere's intrinsic non-Euclidean geometry, leaving a critical gap in rigorous modeling and forecasting techniques. To address this methodological gap, we propose a unified geometric framework to analyze nonstationary spherically embedded time series. Central to our approach is a novel nonparametric spherical trend-periodicity decomposition model that uses an optimal-transport-based removal operation to sequentially extract the smooth trend and periodic components while preserving spherical topology. The resulting de-trended and de-seasonalized stationary residuals can be further modeled using a spherical autoregressive model, formalizing a novel trend-periodic spherical autoregressive model. Theoretical foundations for the modeling procedure are established on the consistency under temporal dependence. Extensive simulations corroborate these theoretical guarantees and demonstrate the superior finite-sample predictive performance of the trend-periodic spherical autoregressive model compared to existing spherical autoregressive models and neural network models. Finally, we validate the practical utility of our methodology through applications to electricity generation compositions and bike trip volume profiles, yielding significantly enhanced forecasting accuracy while providing interpretable insights into the underlying structural dynamics.
\vspace{.1in}

\end{abstract}

\newpage
\spacingset{1.94}

\section{Introduction}

The big data era has seen a surge in the acquisition of non-Euclidean random objects, while these data structures are increasingly prevalent across diverse scientific domains and are frequently observed over time \citep{marron2021object}. A critical class of such time-varying random objects includes time series with values naturally, or can be equivalently, represented as points on a sphere with possibly infinite dimension. This category encompasses directional time series, compositional time series, and distributional time series. We characterize these objects as spherically embedded time series, a framework with significant utility across diverse fields such as energy economics and urban informatics  \citep[see, e.g.,][]{fellner2007new,wagner2017functional}. In the context of energy economics, the analysis of compositional time series of electricity generation allows researchers to track the dynamic structural replacement of carbon-heavy fuels with renewables independently of fluctuations in total electricity demand. Similarly, in urban informatics, characterizing the temporal evolution of daily traffic volume distributions enables researchers to capture latent shape shifts, offering critical insights into urban mobility patterns.

Unlike traditional Euclidean spaces, the intrinsic geometries of directional, compositional, and distributional data do not support standard vector space operations, such as addition or scalar multiplication. This lack of linear structure necessitates the development of statistical methodologies tailored to non-Euclidean manifolds. Recent advances have introduced Wasserstein autoregressive (AR) models for distributional time series (\citealt{ zhang2022wasserstein, chen2023wasserstein, zhu2023autoregressive, ghodrati2024distributional, jiang2026wasserstein}), while a spherical AR model for spherically embedded time series has been developed by \cite{zhu2024spherical}. One of the key advantages of the spherical AR model is that it works for a wider range of object types, such as spherical, compositional, and distributional data, while Wasserstein AR models are limited to distributional time series. 

A significant limitation of the aforementioned methodologies for either distributional time series or a broader class of spherically embedded time series is the assumption of stationarity, either in the original time series or after applying a differencing operator. However, in many empirical applications, spherically embedded time series exhibit non-stationarity driven by unknown deterministic trend and periodic components, and the framework in \cite{zhu2024spherical} lacks the mechanism to identify or decouple these latent structures. Furthermore, while non-stationarity in Euclidean time series can be addressed via simple additive decomposition (i.e., subtraction), such an operator is undefined on the sphere and poses a significant challenge in modeling spherical embedded time series with unknown trend and periodic components.

Motivated by these challenges, we develop a unified geometric framework for the analysis and forecasting of spherically embedded time series with unknown trend and periodic components. At the core of our approach is a novel nonparametric Spherical Trend-Periodicity Decomposition (STPD) model. Driven by a rigorously defined, optimal-transport-based removal operation, this model systematically decouples the spherically embedded time series into a smooth trend, a periodic component, and a stochastic residual process. By further assuming that the resulting de-trended and de-seasonalized residuals follow a stationary spherical AR structure, we formalize the Trend-Periodic Spherical AR (TPSAR) model which is a powerful semiparametric instantiation of our general decomposition framework. To operationalize these models, we propose a comprehensive four-step model fitting and forecasting procedure as follows, where the STPD model is fully realized through the core decomposition workflow in the first two steps, while the complete four-step procedure enables forecasting utilizing the TPSAR model.
\begin{asparaitem}
    \item[Step I:] trend estimation and removal. We estimate the smooth trend using a local Fr\'{e}chet regression (\citealt{petersen2019frechet}), which generalizes the Euclidean local polynomial regression to general metric spaces. Subsequently, we apply the proposed removal operation to systematically extract this estimated trend from the original spherically embedded time series.
    \item[Step II:] periodicity quantification and removal. We identify the unknown period and estimate the periodic component of the de-trended spherically embedded time series. This is achieved by adapting the object-based periodicity quantification techniques of \cite{xu2025quantifying}, which conceptually extend the Euclidean method in \cite{vogt2014nonparametric} to general metric spaces. The removal operation is then applied to extract the estimated periodic component, yielding a stationary, de-trended, and de-seasonalized residual time series.
    \item[Step III:] residual modeling. We apply a spherical AR model to capture the remaining stochastic dependencies of the residual time series.
    \item[Step IV:] prediction. Out-of-sample forecasts are produced by recombining the fitted trend and periodic components with the predictions of the spherical AR residual process, fully leveraging the intrinsic geometry of the TPSAR model.
\end{asparaitem}

\begin{figure}[!htb]
\centering
\includegraphics[width=0.85\linewidth]{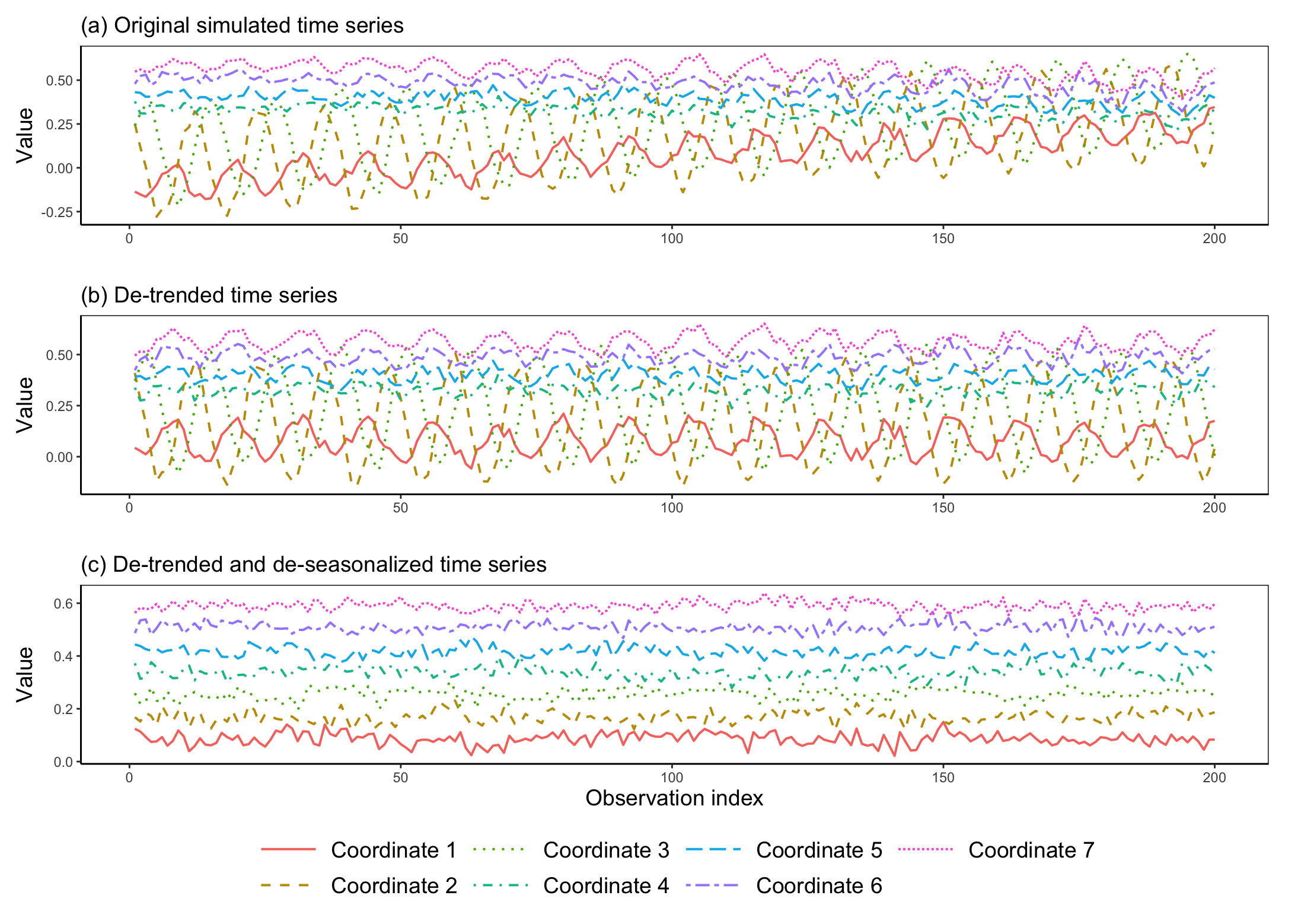}
\caption{The timeline of (a) the simulated spherical time series with unknown trend and periodic components, (b) the de-trended spherical time series, and (c) the de-trended and de-seasonalized residual time series. Individual lines denote the temporal evolution of the spherical vector components, which are constrained to the surface of the sphere.}\label{fig::intro illustrate}
\end{figure}

Figure~\ref{fig::intro illustrate} illustrates the efficacy of the proposed STPD model fitting procedure achieved by the first two steps above using a simulated spherical time series with unknown trend and periodic components. As shown in Figure~\ref{fig::intro illustrate}(a), the original time series exhibits clear nonstationary behavior driven by the trend and periodic components. The trend is effectively neutralized, as evidenced in Figure~\ref{fig::intro illustrate}(b), and the periodic component is further removed in Figure~\ref{fig::intro illustrate}(c), giving a stationary residual time series.

The proposed methodologies constitute a significant advancement in the analysis of spherically embedded time series with unknown trend and periodic components. To our knowledge, this work introduces the first comprehensive framework designed to handle this problem. Beyond its methodological utility, we establish rigorous theoretical foundations for the multi-step estimation procedure. Specifically, we extend the asymptotic results of local Fr\'{e}chet regression (\citealt{petersen2019frechet,chen2022uniform}) and periodicity quantification (\citealt{xu2025quantifying}) which were originally developed under the independence assumption to the more complex case of temporal dependence. This extension provides the necessary theoretical guarantees for consistent estimation in the presence of serial correlation. Extensive simulations showcase the superior accuracy of our method in both estimation and prediction. To validate the empirical utility of our approach, we analyze two real-world datasets on 
\begin{inparaenum}
\item[(i)] monthly U.S. electricity generation compositional time series and \item[(ii)] daily bike trip volume distributional time series in New York City. 
\end{inparaenum}
In both settings, the proposed STPD model yields highly interpretable insights into the latent trend and periodic structures. Notably, the TPSAR model demonstrates superior forecasting accuracy, outperforming the state-of-the-art spherical AR models introduced by \cite{zhu2024spherical} and the deep learning baselines such as Long Short-Term Memory (LSTM) models \citep{yu2019review} and a state-of-the-art fine-tuning transformer-based model, TimePFN \citep{taga2025timepfn}, in both simulations and empirical data analysis.

The remainder of the paper is organized as follows. Section~\ref{sect::spherical ts} introduces the model for spherically embedded time series with trend and periodic components. Section~\ref{sect::modeling procedure} develops the multi-step model fitting procedure and the prediction procedure. Theoretical results, together with the required assumptions, are given in Section~\ref{sect::theory}. Section~\ref{sect::simulation} evaluates the finite-sample performance of the proposed methodologies covering estimation and prediction under a variety of temporal dependence structures. The empirical utility of the proposed framework is demonstrated in Section~\ref{sect::real data} via two real-world case studies. Finally, Section~\ref{sec:conclu} synthesizes our primary contributions. A rolling-window cross-validation criterion for choosing a hyperparameter such as the bandwidth in the local Fr\'{e}chet regression or the AR order in the spherical AR model is provided in the Appendix, and all proofs are deferred to the supplementary material.

\section{Spherically embedded time series with trend and periodic components}\label{sect::spherical ts}

In this section, we first provide a formal definition of spherically embedded time series in Section~\ref{subsect::spherical embedded ts}, and then develop the STPD model for spherically embedded time series with trend and periodic component in Section~\ref{subsect::STPD model}.

\subsection{Spherically embedded time series}\label{subsect::spherical embedded ts}

Spherically embedded time series consist of values that either naturally occur on a sphere or can be represented as elements of the sphere $\mathcal{S}=\{\nu\in\mathcal{H}:\|\nu\|_{\mathcal{H}}=1\}$. In this context, $\mathcal{H}$ represents a separable Hilbert space with the inner product $\langle\cdot,\cdot\rangle_{\mathcal{H}}$ and the norm $\|\nu\|_{\mathcal{H}}=\sqrt{\langle\nu,\nu\rangle_{\mathcal{H}}}$. The metric space $\mathcal{S}$ is equipped with the intrinsic metric $d_\mathcal{S}(\nu_1,\nu_2)=\arccos(\langle\nu_1,\nu_2\rangle_\mathcal{H})$ for $\nu_1,\nu_2\in \mathcal{S}$. This formulation is general to encompass both the finite-dimensional sphere $\mathcal{S}^{m-1}$ where $\mathcal{H}=\mathbb{R}^m$ and the infinite-dimensional Hilbert sphere $\mathcal{S}^{\infty}$.

Such representations are particularly useful for analyzing complex random objects like spherical, compositional, and distributional data. For example, as discussed in \cite{SW11} and \cite{zhu2024spherical}, elements of the $(m-1)$-dimensional simplex $\Delta^{m-1} = \Big\{ (\delta_1,\delta_2,\ldots,\delta_m)^\top \in~\mathbb{R}^m$, $\quad \sum_{l=1}^m \delta_l=1 ~{\rm and}~\delta_l\geq 0,\quad l=1,2,\ldots,m\Big\}$ can be projected onto $\mathcal{S}^{m-1}$ via the pointwise square root transformation $t_p((\delta_1,\delta_2,\ldots,\delta_m)^\top)=(\sqrt{\delta_1},\sqrt{\delta_2},\ldots,\sqrt{\delta_m})^{\top}$ where $^{\top}$ denotes the vector transpose. Similarly, density functions $g$ with support $\mathbb{R}^l$ can be mapped to the infinite-dimensional sphere~$\mathcal{S}^\infty$ using the functional square root transformation $t_f(g) = \sqrt{g(\omega)}$ for all $\omega\in\mathbb{R}^l$. Following the discussion above that a compositional time series or a distributional time series can be equivalently represented as a spherical time series, we mainly focus on developing methodologies for spherical time series with unknown trend and periodic components in the following. 

\subsection{Spherical trend-periodicity decomposition model}\label{subsect::STPD model}

Suppose now that we have a spherical time series $\{y_1,y_2,\ldots,y_T\}\in\mathcal{S}$ and for each $y_t$, it is influenced by a trend component $f(t/T)\in\mathcal{S}$ and a periodic component $g(t)\in\mathcal{S}$ for $t=1,2,\dots,T$. The deterministic trend function $f:[0,1]\mapsto\mathcal{S}$ is assumed to be smooth and the deterministic periodic component $g$ involves a true but unknown integer-valued period $\vartheta_0$ such that $g(l)=g(l+\vartheta_0)=g(l+2\vartheta_0)=\cdots$ for every $l\in\{1,2,\ldots,\vartheta_0\}$. Crucially, the proposed framework remains fully nonparametric with respect to the deterministic structures, where we impose no specific parametric forms on either the trend component $f$ or the periodic component $g$. Moreover, $g$ is treated as a sequence rather than a $\mathcal{S}$-valued function, and our main focus is on the equidistant design, which is the most common situation, as seen in many real-world datasets (\citealt{xu2025quantifying}) and the empirical data examples considered.

\begin{figure}[!htb]
\centering
\includegraphics[width=0.7\linewidth]{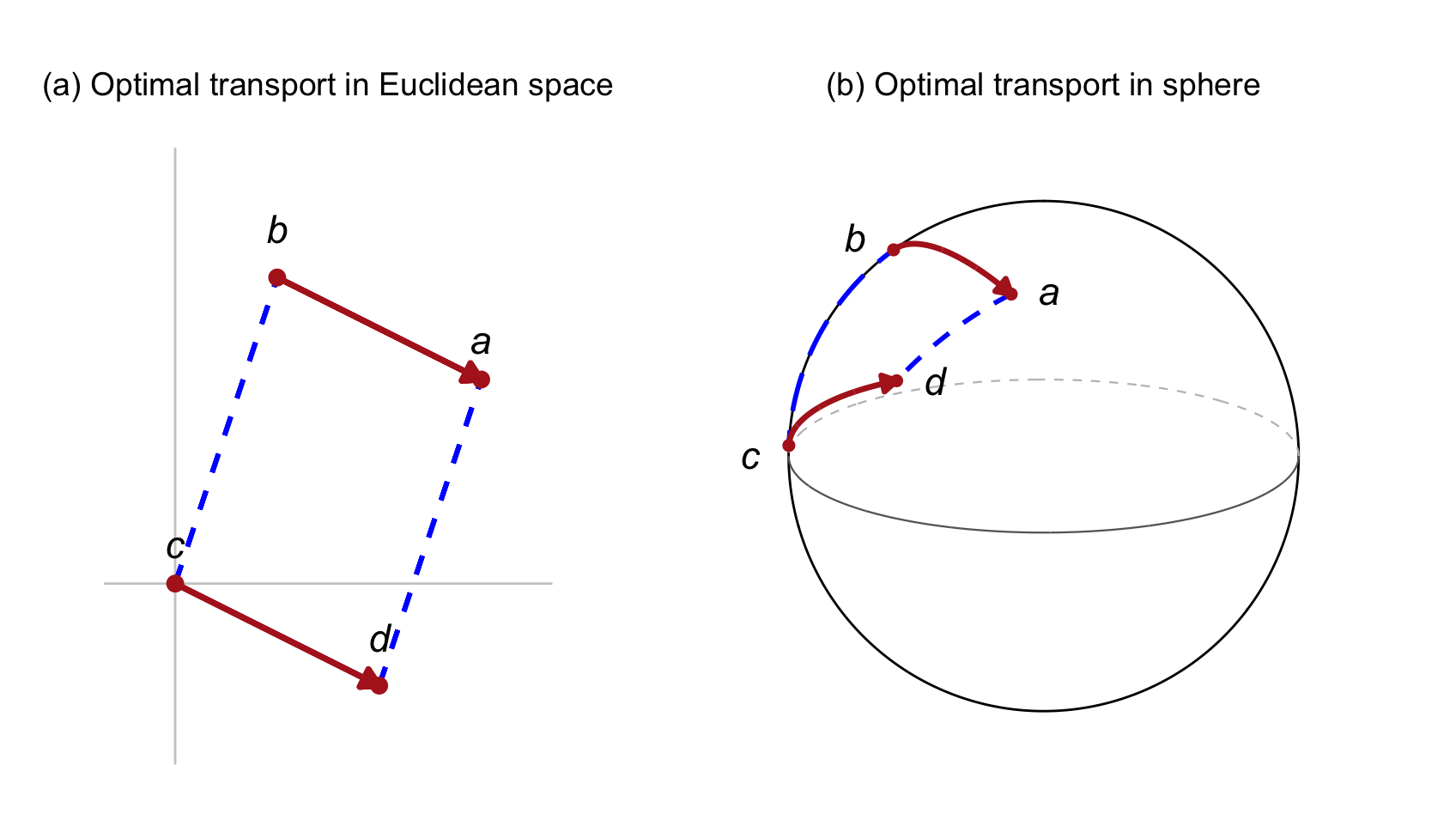}
\caption{Illustration of the removal operation designed to subtract the contribution of $b$ from $a$. This is achieved via: (a) applying the Euclidean optimal transport map $\mathcal{E}_{b\to a}(c)$ with $c=(0,0)^\top$ being a reference origin to obtain $d = a - b + c \in \mathbb{R}^2$; and (b) applying the spherical optimal transport map $\mathcal{M}_{b\to a}(c)$ with $c$ being a reference point to obtain the corresponding point $d \in \mathcal{S}^2$. Solid red lines denote the geodesics of $b \to a$ and $c \to d$, while dashed blue lines indicate the transport of the geodesics between points.}\label{fig::optimal transport visual}
\end{figure}

For a spherical time series influenced by a trend component  and a periodic component, a key challenge here is how to remove the contribution of $f(t/T)$ or $g(t)$ from $y_t$ in a way that ensures that the resulting term remains on the sphere $\mathcal{S}$. To deal with this challenge, we generalize the intuitive notion of a removal operation from Euclidean space to the sphere. In Euclidean space $\mathbb{R}^m$, removing the contribution of~$b$ from~$a$ is equivalent to isolating the relative displacement between them. Given $a, b \in \mathbb{R}^m$ and a reference point $c=0$, this removal is performed by the operation $d = a - b + c$. Mathematically, this removal is based on an optimal transport map that can transport $c$ to the unique point $d$, defined as $\mathcal{E}_{b\to a}(c) =a+c - b$. One can view the distance between $b$ and $a$ as the velocity required to move $b$ to $a$, and applying that same velocity to $c$ gives $d$. As illustrated in Figure~\ref{fig::optimal transport visual}(a), this removal operation $\mathcal{E}_{b\to a}(c)$ transports the entire geodesic between $b$ and $a$ to a new geodesic between $c$ and $d$. Because Euclidean space is flat, this transport is a simple linear translation.

Extending this geometric interpretation of the removal operation in Euclidean space, for $\nu_1,\nu_2\in\mathcal{S}$, a removal operation that removes the contribution of $\nu_2$ from $\nu_1$ can be constructed via a suitable optimal transport map given a reference point $\nu_3$ in the sphere. We denote such a removal operation as $\mathcal{M}_{\nu_2 \to \nu_1}(\nu_3)$. Here we consider adapting an optimal transport  map for spherical data proposed by \cite{zhu2024spherical, zhu2025geodesic}, this gives $\mathcal{M}_{\nu_2 \to \nu_1}(\nu_3)={\rm Exp}(T_{\nu_2,\nu_3})\nu_1$ where ${\rm Exp}$ is the exponential map and $T_{\nu_2,\nu_3}$ is a skew-symmetric operator given as
\begin{align}\label{formula::optimal transport}
T_{\nu_2,\nu_3}=\nu_3 \ominus \nu_2 = \eta (\zeta_2\otimes \zeta_1 - \zeta_1\otimes\zeta_2)
\end{align}
with $\eta=d_\mathcal{S}(\nu_2,\nu_3)$, $\zeta_1=\nu_2$, $\zeta_2=(\nu_3-\langle\nu_3,\nu_2\rangle_{\mathcal{H}}\nu_2)/\|\nu_3-\langle\nu_3,\nu_2\rangle_{\mathcal{H}}\nu_2\|_{\mathcal{H}}$, and $\otimes$ denoting the tensor product.  The optimal transport ${\rm Exp}(T_{\nu_2,\nu_3})$ is a rotation operator that rotates the sphere within ${\rm span}\{\nu_2,\nu_3\}$ by the angle $\eta$, and has an explicit form as 
\begin{align}\label{formula::removal operator}
{\rm Exp}(T_{\nu_2,\nu_3})= {\rm id}  + \frac{\sin(\eta)}{\eta}T_{\nu_2,\nu_3} + \frac{1-\cos(\eta)}{\eta^2}T_{\nu_2,\nu_3}^2,
\end{align}
where ${\rm id}$ denotes the identity operator. The above construction ensures that ${\rm Exp}(T_{\nu_2,\nu_3})\nu_2=\nu_3$, meaning it can move $\nu_2$ to $\nu_3$. As illustrated in Figure~\ref{fig::optimal transport visual}(b), for $a,b\in\mathcal{S}$ and a reference point $c\in\mathcal{S}$, the removal operation $\mathcal{M}_{b \to a}(c)$ applies the velocity required to move $b$ to $a$ to the reference point $c$, giving a unique point $d\in\mathcal{S}$ and ensuring that the geodesics of $b \to a$ and $c \to d$ are the same, similar to the case in the Euclidean space in Figure~\ref{fig::optimal transport visual}(a). Notably, while the removal operation $\mathcal{M}_{b \to a}(c)$ is constructed in an intrinsic way, Section~S2 in the supplementary material discusses the failure when we consider directly using some simple extrinsic ways for spherical data and compositional data.

By choosing the reference point $\mu_Y$ as the overall population Fr\'{e}chet mean 
\begin{align}\label{formula:: overall Frechet mean}
\mu_Y=\argmin_{\nu\in\mathcal{S}}\mathbb{E}\left\{T^{-1}\sum_{t=1}^T d_\mathcal{S}^2(\nu,y_t)\right\},
\end{align}
one can then remove the contribution of $f(t/T)$ from $y_t$ given the reference point $\mu_Y$ as 
\begin{align}\label{map1::remove trend}
R_t^{(1)} = \mathcal{M}_{f(t/T)\to y_t}(\mu_Y).
\end{align}
Thus, $\mathcal{M}_{f(t/T)\to y_t}\mu_Y$ in~\eqref{map1::remove trend} can be interpreted as transporting the relative gap between $f(t/T)$ and $y_t$ to a reference point $\mu_Y$, similar to the de-trending step in real-valued time series. Notably, the choice of reference point is vital as different reference points will lead to different points after the removal. Here, the reason we select $\mu_Y$ as the reference point is that, if $\{y_t\}_{i=1}^T$ is only influenced by a trend $f$ which takes a constant value, then we have $f(u)=\mu_Y$ for every $u\in[0,1]$, and (\ref{map1::remove trend}) ensures that $R_t^{(1)}=y_t$ for every $t\in \{1,2,\ldots,T\}$. This means, if the spherical time series does not have a non-constant trend, then the de-trended spherical time series $\left\{R_t^{(1)}\right\}_{t=1}^T$ in (\ref{map1::remove trend}) is just the original time series $\{y_t\}_{t=1}^T$.

Suppose that once the contribution of the trend component $f$ is removed, the de-trended spherical time series $\left\{R_t^{(1)}\right\}_{t=1}^T$ is driven solely by the periodic component $g$.  This gives 
\begin{equation}\label{formula::periodic component}
g(r(t,\vartheta_0)) = \argmin_{\nu\in\mathcal{S}} \mathbb{E} \left\{ d_\mathcal{S}^2\left(\nu,R_t^{(1)}\right) \right\},    
\end{equation}
where $r(t,\vartheta_0)=t+\vartheta_0-\vartheta_0\lfloor (t+\vartheta_0-1)/\vartheta_0 \rfloor$ with $\lfloor\cdot\rfloor$ being the floor function. Note that we can further remove the periodic term $g(t)$ from $R_t^{(1)}$ as
\begin{align}\label{map2::remove periodic term}
    R_t^{(2)} =  \mathcal{M}_{g(t)\to R_t^{(1)}}(\mu_{R,1})
\end{align}
where $\mu_{R,1}=\argmin_{\nu\in\mathcal{S}}\mathbb{E}\left\{T^{-1}\sum_{t=1}^T d_\mathcal{S}^2\left(\nu,R_t^{(1)}\right)\right\}$. Similar to the discussion on the reference point choice in~\eqref{map1::remove trend}. If the true period $\vartheta_0=1$, meaning there is no periodic structure in the de-trended time series, then one always has $g(t)=\mu_{R,1}$, leading to $R_t^{(2)}=R_t^{(1)}$ for every $t\in \{1,2,\ldots,T\}$. This indicates that, if the time series $\left\{R_t^{(1)}\right\}_{t=1}^T$ is not influenced by any periodic component, then the removal of the periodicity in (\ref{map2::remove periodic term}) will remove nothing.

In summary, for a spherical time series $\{y_1,y_2,\ldots,y_T\}$ influenced by a trend component $f(t/T)$ and a periodic component $g(t)$, one could assume that it follows the following STPD model by combining (\ref{map1::remove trend}) and (\ref{map2::remove periodic term})
\begin{align}\label{final model}
    R_t^{(2)} = \mathcal{M}_{g(t)\to R_t^{(1)}}(\mu_{R,1}), \quad ~~ R_t^{(1)} = \mathcal{M}_{f(t/T)\to y_t}(\mu_Y)
\end{align}
where $\left\{R_t^{(2)}\right\}_{t=1}^T$ is assumed to be strictly stationary with weak temporal dependence. 

It is important to note that, unlike the Euclidean case where the order of component removal is arbitrary, the sequence of removal operations is critical for spherically embedded time series. This arises from the intrinsic geometry of the sphere and the definition of the removal operation $\mathcal{M}_{\nu_2\to\nu_1}(\nu_3)$ in (\ref{formula::removal operator}). Consequently, the STPD model (\ref{final model}) necessitates a specific multi-step procedure where first estimating the smooth trend $f$, followed by estimating the periodic component $g$. 

The natural question then arises: can this order be reversed to estimate the periodic component before the trend? The answer is negative. To see why, suppose that the removal order in the STPD model~\eqref{final model} was switched, one would then require $g(t)= \argmin_{\nu\in\mathcal{S}} \mathbb{E} \{ d_\mathcal{S}^2(\nu,y_t) \}$ to achieve the consistent estimation of the periodic component, as seen in Section~\ref{sect::component est procedure}. However, since the periodic component $g$ possesses a global structure such that $g(l) = g(l + k\vartheta_0)$ for $k \in \{1,2,3,\ldots\}$, the presence of a non-constant trend in $\{y_t\}_{t=1}^T$ leads to the requirement that the relationship $g(l) = g(l + k\vartheta_0)$ cannot be satisfied.

\section{Modeling and prediction procedure}\label{sect::modeling procedure}

In this section, we first develop the multi-step modeling procedure covering the estimation of the trend and periodic components in Section~\ref{sect::component est procedure} and a model fitting procedure for the final residuals $\left\{R_t^{(2)}\right\}_{t=1}^T$ in Section~\ref{sect::AR fitting}. Built upon the multi-step modeling procedure, the prediction procedure is then presented in Section~\ref{sect::prediction}.

\subsection{Trend and periodic component estimation procedure}\label{sect::component est procedure}

The estimation procedure for the trend and periodic components in the STPD model involves three nonparametric estimation steps: 
\begin{inparaenum}
\item[(i)] estimate the unknown trend $f$ in Section~\ref{subsect::trend estimation}, 
\item[(ii)] given the estimated trend, estimate the unknown period $\vartheta_0$ in Section~\ref{sect::period est}, and 
\item[(iii)] given the estimated trend and period, estimate the periodic component $g$ in Section~\ref{sect::periodic component est}.
\end{inparaenum}

\subsubsection{Estimation of the trend component}\label{subsect::trend estimation}

Note that the trend is the low-frequency component of the time series $\{y_t\}_{t=1}^T$, allowing us to utilize a local Fr\'{e}chet regression (\citealt{petersen2019frechet}) that extends the local polynomial regression in the Euclidean space to a general metric space.  The local Fr\'{e}chet regression estimator takes the form of
\begin{align}\label{formula::trend estimator}
\widehat{f}(u)=\argmin_{\nu\in\mathcal{S}} \widehat{F}(\nu,u),~~\widehat{F}(\nu,u)= T^{-1}\sum_{t=1}^T \widehat{w}(t/T,u) d_\mathcal{S}^2(\nu,y_t)
\end{align}
for $u\in[0,1]$, where $\widehat{w}(t/T,u)=\widehat{\sigma}^{-2}K_h(t/T-u)[\widehat{\tau}_2(u)-\widehat{\tau}_1(u)(t/T-u)]$ with $\widehat{\tau}_l(u)=T^{-1}\sum_{t=1}^T[K_h(t/T-u)(t/T-u)^l]$ for $l\in\{0,1,2\}$, and $\widehat{\sigma}^2=\widehat{\tau}_0(u)\widehat{\tau}_2(u)-\widehat{\tau}_1^2(u)$. In practice, we consider using the Gaussian kernel and the bandwidth $h$ is selected via a rolling-window cross validation, whose details are given in the Appendix. Although the trend estimator $\widehat{f}$ in (\ref{formula::trend estimator}) is a local estimator governed by a bandwidth parameter $h$, it is expected to smooth out the periodic component rather than absorbing it, both theoretically and in practice. In practice, an extremely small $h$ is unlikely to be selected by rolling-window cross-validation due to the bias-variance trade-off. Consequently, as demonstrated in our numerical studies, $\widehat{f}$ isolates the smooth trend without being influenced by the periodicity. Thus, the periodic structure remains in the de-trended time series and must be evaluated subsequently. Most importantly, from a theoretical perspective, as the sample size $T\to\infty$, the number of periods captured within the local smoothing window diverges to infinity. This ensures that the periodic fluctuations average out asymptotically, a property we rigorously verify in Section~\ref{subsect::trend and periodic component est theory}.

\subsubsection{Estimation of the period}\label{sect::period est}

Given the estimated trend component $\widehat{f}$ in (\ref{formula::trend estimator}), calculate $\widehat{R}_t^{(1)} = \mathcal{M}_{\widehat{f}(t/T)\to y_t}\left(\widehat{\mu}_Y\right)$ for each $t=1,2,\ldots,T$ where $\widehat{\mu}_Y=\argmin_{\nu\in\mathcal{S}}T^{-1}\sum_{t=1}^T d_\mathcal{S}^2(\nu,y_t)$. Following the methodology developed by \cite{xu2025quantifying}, period estimation can be framed as a model selection problem as follows.  

Consider a set of candidate periods $\vartheta \in \{1, 2, \ldots, \Theta_T\}$, where $\Theta_T$ represents the maximum allowable period length. For any given candidate period $\vartheta$, we derive an appropriate model $\widehat{g}_t(\vartheta, T)$, $t=1,2,\ldots,T$, and compute the residual sum of squares (RSS) ${\rm RSS}(\vartheta) = \sum_{t=1}^T d_\mathcal{S}^2\left(\widehat{R}_t^{(1)} , \widehat{g}_t(\vartheta, T) \right)$. Ideally, the RSS behaves according to the following three regimes with high probability:
\begin{asparaitem}
    \item (Correct Specification) When $\vartheta$ matches the true period $\vartheta_0$, ${\rm RSS}(\vartheta)$ is small.
    \item (Overfitting) For any $\vartheta = k\vartheta_0$ for any positive integer $k>1$, the model $\widehat{g}_t(\vartheta, T)$ overfits, resulting in ${\rm RSS}(\vartheta) < {\rm RSS}(\vartheta_0)$.
    \item (Misspecification) For any $\vartheta$ that is not a multiple of $\vartheta_0$, the model $\widehat{g}_t(\vartheta, T)$ is poorly fitted, so that ${\rm RSS}(\vartheta) > {\rm RSS}(\vartheta_0)$.
\end{asparaitem}
To account for the overfitting at higher multiples of the true period, we define a penalized RSS as 
\begin{equation}\label{formula::loss function}
\mathcal{L}(\vartheta,\lambda_T) = {\rm RSS}(\vartheta) + \lambda_T \vartheta
\end{equation}
The estimated period $\widehat{\vartheta}_{\lambda_T}$ is then the candidate that minimises this objective as
\begin{equation}\label{formula::period estimator}
\widehat{\vartheta}_{\lambda_T} = \argmin_{1\leq \vartheta \leq \Theta_T} \mathcal{L}(\vartheta,\lambda_T).
\end{equation}
The tuning parameter $\lambda_T$ is designed to diverge to infinity at an appropriate rate to ensure that the penalty $\lambda_T \vartheta$ effectively suppresses the overfitting situation; see Theorem \ref{thm::period estimation consistency} for the formal consistency requirements.

As seen above, the key to the period estimation is the construction of the suitable model $\widehat{g}_t(\vartheta, T)$. Following \cite{xu2025quantifying}, this can be achieved through a global Fr\'{e}chet regression model (\citealt{petersen2019frechet}) given by
\begin{align}\label{formula::period est model}
\widehat{g}_t(\vartheta, T) = \argmin_{\nu\in\mathcal{S}}  \widehat{G}_t(\nu; \vartheta, T),~~
\widehat{G}_t(\nu; \vartheta, T) = T^{-1}\sum_{i=1}^T s^{(t)}_i d_\mathcal{S}^2\left(\nu,\widehat{R}_i^{(1)}\right),
\end{align}
for $t=1,\ldots,T$, where $s^{(t)}_i$ is constructed as the $i$\textsuperscript{th} column of $x_t^\top(X^\top X)^{-1}X^\top$ with $x_t=x_t(\vartheta)=(x_{t1},\ldots,x_{t\vartheta})^\top$, $x_{tj}=1$ for $j=t+\vartheta-\vartheta\lfloor (t+\vartheta-1)/\vartheta \rfloor$ and $x_{tk}=0$ for $k\neq j$, and $X=X(\vartheta)=(x_1,\ldots,x_T)^\top$. Observe that specifying $\vartheta$ and sample size $T$ completely determines both $x_t$ and $X$. Consequently, the weights $\left\{s^{(t)}_i\right\}_{i=1}^T$ associated with $G_t(\nu; \vartheta, T)$ are strictly a function of $\vartheta$ and $T$. Moreover, $\widehat{g}_t(\vartheta_0, T)$ in (\ref{formula::period est model}) is constructed as a consistent estimator of the periodic component $g(r(t,\vartheta_0))$ in (\ref{formula::periodic component}), as seen in Lemma~S1 in the supplementary material.

In particular, the maximum allowable period length $\Theta_T$ is permitted to grow with the sample size $T$ at an appropriate rate, as established in Theorem~\ref{thm::period estimation consistency}. This flexibility is particularly useful in practice when prior knowledge of the search space is unavailable. In such cases, $\Theta_T$ can be determined as a function of the sample size $T$.

Within the penalized RSS in~\eqref{formula::loss function}, the tuning parameter $\lambda_T$ governs a critical trade-off, as a small $\lambda_T$ may lead to overfitting, while a large $\lambda_T$ could cause misspecification. Both cases ultimately affect the consistent recovery of the underlying true period $\vartheta_0$. To resolve this issue, we deploy a general regularisation function $\ell(T)$ to define an information criterion of the form
\begin{align} \label{eq::BIC}
 {\rm IC}_\lambda = \log \{ {\rm RSS}(\widehat{\vartheta}_\lambda) / T \} + \widehat{\vartheta}_\lambda \ell(T)
\end{align}
can consistently estimate the true period $\vartheta_0$. Using (\ref{eq::BIC}), the tuning parameter $\lambda$ is selected as 
\begin{align}\label{formula::lambda selection}
    \widehat{\lambda}=\argmin_\lambda {\rm IC}_\lambda.
\end{align}
Using the setting $\lambda_T=\widehat{\lambda}$ in (\ref{formula::period estimator}), the final period estimator is then $\widehat{\vartheta}_{\widehat{\lambda}}$.

\subsubsection{Estimation of the periodic component}\label{sect::periodic component est}

Given the final period estimator $\widehat{\vartheta}_{\widehat{\lambda}}$ and letting $\mathbb{N}^+$ be the set of positive integers, we can subsequently estimate the entire periodic sequence $\{g(t)\}_{t\in\mathbb{N}^+}$ defined in (\ref{formula::periodic component}) as
\begin{align}\label{formula::periodic component estimator}
    \widehat{g}(l)=\widehat{g}_l\left(\widehat{\vartheta}_{\widehat{\lambda}},T\right),
\end{align}
and $\widehat{g}(l+k\widehat{\vartheta}_{\widehat{\lambda}}) = \widehat{g}(l)$ for $l=1,2,\ldots,\widehat{\vartheta}_{\widehat{\lambda}}$ and all $k=0,1,2,\ldots$. This guarantees that $\widehat{g}_{l+k\widehat{\vartheta}_{\widehat{\lambda}}}\left(\widehat{\vartheta}_{\widehat{\lambda}},T\right)=\widehat{g}_{l}\left(\widehat{\vartheta}_{\widehat{\lambda}},T\right)$ holds for any $l=1,2,\ldots,\widehat{\vartheta}_{\widehat{\lambda}}$ and non-negative integer $k$. Consequently, the estimated sequence $\{\widehat{g}(t)\}_{t=1}^T$ strictly preserves a periodic structure with period $\widehat{\vartheta}_{\widehat{\lambda}}$.

\subsection{Model fitting procedure for residual time series}\label{sect::AR fitting}

Recall that in (\ref{final model}) the final residual time series $\left\{R_t^{(2)}\right\}_{t=1}^T$ is assumed to be strictly stationary with weak temporal dependence; thus, one can use a suitable spherical AR model to model the final residual time series where the model setup is introduced in Section~\ref{sect::spherrical AR setup} and the AR coefficient estimation is provided in Section~\ref{sect::AR coef est}.

\subsubsection{Trend-periodic spherical AR model setup}\label{sect::spherrical AR setup}

First note that for real-valued $a$ and $b$, $a-b$ can be interpreted as a generator of the transformation that moves $b$ to $a$ along a geodesic in Euclidean space. For spherical data $\nu_1,\nu_2\in \mathcal{S}$, there also exists a generator of such a transformation that  moves $\nu_2$ to $\nu_1$ along a geodesic by a rotation given in (\ref{formula::optimal transport}) and taking the form of $\nu_1\ominus\nu_2$. The generator $\nu_1\ominus\nu_2$ lies in a set of skew-symmetric operators, denoted $\mathcal{C}(\mathcal{H})$, which is Hilbertian with the Hilbert-Schmidt inner product as $\langle h_1\otimes \widetilde{h}_1, h_2\otimes \widetilde{h}_2 \rangle_{\mathcal{C}(\mathcal{H})} = \langle h_1 , h_2 \rangle_{\mathcal{H}} \langle \widetilde{h}_1 , \widetilde{h}_2 \rangle_{\mathcal{H}}$.  Then, a suitable spherical AR model of order $p$ for the stationary residual time series $\left\{R_t^{(2)}\right\}_{t=1}^T$ can be the one proposed by \cite{zhu2024spherical} as
\begin{align}\label{model::spheircal AR model}
    \Xi_t - \mu_{\Xi} = \varphi_1 (\Xi_{t-1} - \mu_{\Xi}) + \cdots + \varphi_p (\Xi_{t-p} - \mu_{\Xi}) + \epsilon_t
\end{align}
where $\Xi_t=R_t^{(2)} \ominus \mu_{R,2}$, $\mu_{R,2}=\argmin_{\nu\in\mathcal{S}}\mathbb{E}\left[d_\mathcal{S}^2\left(\nu,R_1^{(2)}\right)\right]$, $\mu_{\Xi}=\mathbb{E}(\Xi_t)$, $\varphi_1,\varphi_2,\ldots,\varphi_p\in\mathbb{R}$ are the spherical AR coefficients, and $\{\epsilon_t\}_{t=1}^T\subset \mathcal{C}(\mathcal{H})$ are independent and identically distributed random noise with mean 0. Note that the model~\eqref{model::spheircal AR model} works for both finite- and infinite-dimensional spheres. By integrating the decomposition of deterministic trend and periodic structures in~\eqref{final model} with the spherical AR modeling of the final stationary residuals in~\eqref{model::spheircal AR model}, we define this unified semiparametric framework as the TPSAR model

Following \citet[][Theorem~2]{zhu2024spherical}, when $\mathbb{E}(\langle \epsilon_t,\epsilon_t \rangle_{\mathcal{C}(\mathcal{H})})<\infty$ and the roots of the characteriztic function $a(z)=1-\varphi_1z-\varphi_2z^2-\cdots-\varphi_pz^p$ are outside the unit circle, then $\Xi_t-\mu_{\Xi}=\sum_{i=0}^\infty \iota_i\epsilon_{t-i}$ is a unique stationary solution of (\ref{model::spheircal AR model}) where $\{\iota_t\}_{t=0}^\infty$ is absolutely summable and determined by $1/a(z)=\sum_{i=0}^\infty \iota_i z^i$.

\subsubsection{Parameter estimation}\label{sect::AR coef est}

Similarly to the estimation procedure in \cite{zhu2024spherical}, one can use Yule--Walker type estimators for the AR coefficients $\varphi_1,\varphi_2,\ldots,\varphi_p$ of the model~(\ref{model::spheircal AR model}). Let $\varrho_k=\mathbb{E}(\langle \Xi_1 - \mu_{\Xi},\Xi_{k+1} - \mu_{\Xi} \rangle_{\mathcal{C}(\mathcal{H})})$ for $k=0,1,\ldots,p$ and the AR model parameters in~(\ref{model::spheircal AR model}) satisfy 
\[
\begin{pmatrix}
    \varrho_1\\
    \varrho_2\\
    \vdots\\
    \varrho_p
\end{pmatrix} = \begin{pmatrix}
    \varrho_0 & \varrho_1 & \cdots &\varrho_{p-1} \\
    \varrho_1 & \varrho_0 & \cdots &\varrho_{p-2} \\
    \vdots & \vdots && \vdots \\
    \varrho_{p-1} & \varrho_{p-2} & \cdots &\varrho_0 \\
\end{pmatrix} \begin{pmatrix}
    \varphi_1\\
    \varphi_2\\
    \vdots\\
    \varphi_p
\end{pmatrix}.
\]
Then, after getting all the estimations $\widehat{\vartheta}_{\widehat{\lambda}},\widehat{g},\widehat{f}$, and $\widehat{\mu}_Y$, and plugging them into~(\ref{final model}), we have $\widehat{\Xi}_t=\widehat{R}_t^{(2)} \ominus \widehat{\mu}_{R,2}$ where $\widehat{\mu}_{R,2}=\argmin_{\nu\in\mathcal{S}}T^{-1}\sum_{t=1}^T d_\mathcal{S}^2\left(\nu,\widehat{R}_t^{(2)}\right)$, $\widehat{R}_t^{(2)} = \mathcal{M}_{\widehat{g}(t)\to \widehat{R}_t^{(1)}}\left(\widehat{\mu}_{R,1}\right),  \widehat{R}_t^{(1)} = \mathcal{M}_{\widehat{f}(t/T)\to y_t}\left(\widehat{\mu}_Y\right)$, and $\widehat{\mu}_{R,1}=\argmin_{\nu\in\mathcal{S}}T^{-1}\sum_{t=1}^T d_\mathcal{S}^2\left(\nu,\widehat{R}_t^{(1)}\right)$. This allows us to obtain the AR coefficient estimation as
\begin{align}\label{formula::AR coef estimation}
\begin{pmatrix}
    \widehat{\varphi}_1\\
    \widehat{\varphi}_2\\
    \vdots\\
    \widehat{\varphi}_p
\end{pmatrix} = \begin{pmatrix}
    \widehat{\varrho}_0 & \widehat{\varrho}_1 & \cdots &\widehat{\varrho}_{p-1} \\
    \widehat{\varrho}_1 & \widehat{\varrho}_0 & \cdots &\widehat{\varrho}_{p-2} \\
    \vdots & \vdots && \vdots \\
    \widehat{\varrho}_{p-1} & \widehat{\varrho}_{p-2} & \cdots &\widehat{\varrho}_0 \\
\end{pmatrix}^{-1} \begin{pmatrix}
    \widehat{\varrho}_1\\
    \widehat{\varrho}_2\\
    \vdots\\
    \widehat{\varrho}_p
\end{pmatrix},    
\end{align}
where $\widehat{\varrho}_k = (T-k)^{-1}\sum_{i=1}^{T-k} \left\langle \widehat{\Xi}_i - \widehat{\mu}_\Xi,  \widehat{\Xi}_{i+k} - \widehat{\mu}_\Xi  \right\rangle_{\mathcal{C}(\mathcal{H})}$ and $\widehat{\mu}_\Xi=T^{-1}\sum_{t=1}^T \widehat{\Xi}_t$.
In practice, one needs to select the spherical AR order $p$, which can be done via a rolling-window cross-validation approach given in the Appendix.

\subsection{Prediction procedure}\label{sect::prediction}

Given a training sample $y_1,y_2,\ldots,y_T$, suppose we have all the required estimations obtained from Sections~\ref{sect::component est procedure} and \ref{sect::AR fitting} as $\widehat{f}$ for the trend component, $\widehat{g}$ for the periodic component with $\widehat{\vartheta}_{\widehat{\lambda}}$ for the period, $\widehat{\varphi}_1,\ldots,\widehat{\varphi}_p$ for the spherical AR coefficients with the selected order $p$, $\widehat{\mu}_\Xi$ for the sample mean of $\left\{\widehat{\Xi}_t\right\}_{t=1}^T$, $\widehat{\mu}_{R,2}$ for the Fr\'{e}chet mean of $\left\{\widehat{R}_t^{(2)}\right\}_{t=1}^T$, $\widehat{\mu}_{R,1}$ for the Fr\'{e}chet mean of $\left\{\widehat{R}_t^{(1)}\right\}_{t=1}^T$, and $\widehat{\mu}_Y$ for the Fr\'{e}chet mean of $\{y_t\}_{t=1}^T$. Then, the $m$-step-ahead prediction $\widehat{y}_{T+m|T}$ can be obtained as 
\begin{align*}
    \widehat{y}_{T+m|T}=\mathcal{M}_{\widehat{\mu}_Y\to \widehat{R}_{T+m|T}^{(1)}} \left[\widehat{f}\Big(\frac{T+m}{T}\Big)\right],
\end{align*}
where $\widehat{R}_{T+m|T}^{(1)}=\mathcal{M}_{\widehat{\mu}_{R,1}\to \widehat{R}_{T+m|T}^{(2)}} \left( \widehat{g}(T+m)\right)$, $\widehat{R}_{T+m|T}^{(2)}={\rm Exp}\left(\widehat{\Xi}_{T+m|T}\right)\widehat{\mu}_{R,2}$, and $\widehat{\Xi}_{T+m|T} = \widehat{\mu}_\Xi + \sum_{l=1}^p \widehat{\varphi}_l (\widetilde{\Xi}_{T+m-l} - \widehat{\mu}_\Xi)$ with $\widetilde{\Xi}_{T+m-l} =\widehat{\Xi}_{T+m-l}$ if $m-l \leq 0$, and $\widetilde{\Xi}_{T+m-l}=\widehat{\Xi}_{T+m-l|T}$ otherwise.

\section{Theoretical results}\label{sect::theory}

This section first presents the technical assumptions in Section~\ref{subsect::assumption}, and then provides the theoretical support for the trend and periodic component estimation in Section~\ref{subsect::trend and periodic component est theory} and the estimation of spherical AR parameters in Section~\ref{subsect::AR coef est theory}.

\subsection{Preliminaries and Assumptions}\label{subsect::assumption}

Before we list the required assumptions, we first revisit the $\alpha$-mixing condition for spherically embedded time series following \cite{zhang2025change}. For spherical random variables $\{y_t\}_{t\in\mathbb{Z}}$, denote the $\sigma$-field generated by $\{y_{t_1},y_{t_1+1},\ldots,y_{t_2}\}$ with $-\infty\leq t_1\leq t_2 \leq \infty$ as $\mathcal{F}_{t_1}^{t_2}$. The sequence $\{y_t\}_{t\in\mathbb{Z}}$ is said to be $\alpha$-mixing if the $\alpha$-mixing coefficients $\alpha(n) \to 0$ as $n \to \infty$, where $\alpha(n)=\sup_{\mathcal{A}\in\mathcal{F}_{-\infty}^0,\mathcal{B}\in\mathcal{F}_n^{\infty}}\left| \mathbb{P}(\mathcal{A}\cap \mathcal{B})-\mathbb{P}(\mathcal{A})\mathbb{P}(\mathcal{B}) \right|$. The sphere $\mathcal{S}$ is of negative type (\citealt{lyons2020strong}) and by \cite{schoenberg1937certain,schoenberg1938metric}, there exist a Hilbert space $\mathcal{H}$ and an embedding map $\phi:\mathcal{S}\mapsto \mathcal{H}$ such that $d_\mathcal{S}(y,y^\prime)=\| \phi(y) - \phi(y^\prime) \|_\mathcal{H}^2$. For $t\in\mathbb{Z}$, let $z_t=\phi(y_t)$, and one should notice that $\phi(\cdot)$ is a continuous bijection between $(\mathcal{S},d_\mathcal{S})$ and $(\phi(\mathcal{S}),d_\mathcal{H})$ with continuous inverse. Then, for any positive integers $t_1\leq t_2$, the $\sigma$-algebra generated by $\{y_{t_1},y_{t_1+1},\ldots,y_{t_2}\}$ is the same as the one generated by $\{z_{t_1},z_{t_1+1},\ldots,z_{t_2}\}$. This means that if $\{y_t\}_{t\in\mathbb{Z}}$ is $\alpha$-mixing, then one always has $\{z_t\}_{t\in\mathbb{Z}}$ is $\alpha$-mixing with the same mixing coefficient. Following similar ideas, one can observe that, when the spherical AR model (\ref{model::spheircal AR model}) has unique stationary solution as discussed in Section~\ref{sect::spherrical AR setup}, then $\{y_t\}_{t\in\mathbb{Z}}$ is $\alpha$-mixing with the mixing coefficients always satisfy the polynomial decay rate listed in Assumption~\ref{assump::4} below.

For the time series $\{y_t\}_{t=1}^T$ influenced by a smooth trend and a periodic component, consider $\mathcal{L}=\{0,1,\ldots,\vartheta_0-1\}$ and distributions $\{P_{u,l}\}_{u\in[0,1],l\in\mathcal{L}}$. Assume for each $t\in\{1,2,\ldots,T\}$, $y_t$ follows a distribution $P_{u_t,l_t}$ where $u_t=t/T$ and $l_t=t~{\rm mod}~\vartheta_0$. Here ${\rm mod}$ is the modulo operation. For $u\in[0,1]$ and $l\in\mathcal{L}$, define $Q(\nu,u,l)=\mathbb{E}_{Y\sim P_{u,l}}\left[ d_\mathcal{S}^2(\nu,Y)\right]$ and let 
\begin{align}\label{formula::trend component}
    f(u)=\argmin_{\nu\in\mathcal{S}}F(\nu,u),~~F(\nu,u)=\vartheta_0^{-1}\sum_{l=0}^{\vartheta_0-1}Q(\nu,u,l).
\end{align}
In Section~S3 in the supplementary material, we show that (\ref{formula::trend component}) reduces to the classical setting of the real-valued time series with trend and periodic components when replacing the sphere to the Euclidean space.

The technical assumptions are listed as follows.

\begin{assumption}\label{assump::1}
The support ${\rm supp}(Y)\subset \mathcal{S}$ of $\{y_t\}_{t=1}^T$ satisfies $\sup_{\nu_1,\nu_2\in{\rm supp}(Y)}d_\mathcal{S}(\nu_1,\nu_2)\leq \pi/2$. Further assume that for every $t=1,2,\ldots,T$, $d_\mathcal{S}(y_t,f(t/T))\leq \pi/4$ and $d_\mathcal{S}\left(R_t^{(1)},g(t)\right)\leq \pi/4$.
\end{assumption}

\begin{assumption}\label{assump::2}
Let the kernel $K$ be a symmetric probability density function on $\mathbb{R}$ that is uniformly continuous. For for $l_1,l_2\in\{1,2,4,6\}$, define $H_{l_1l_2}=\int_\mathbb{R} K^{l_1}(x)x^{l_2}dx<\infty$, and $H_{14}$ and $H_{26}$ are bounded. Assume that the derivative $K^\prime$ exists and is bounded over the support of $K$, i.e., $\sup_{K(x)>0}|K^\prime(x)|<\infty$. Moreover, require that $\int_\mathbb{R}x^2|K^\prime(x)|\sqrt{|x\log |x||}dx<\infty$.
\end{assumption}

\begin{assumption}\label{assump::3}
For each $l \in \mathcal{L}$, the density function $f_{u,l}(y)$ of the distribution $P_{u,l}$ is twice continuously differentiable with respect to $u \in [0, 1]$ for all $y \in \mathcal{S}$. Furthermore, its second derivative is uniformly bounded as $\sup_{u \in [0,1]} \sup_{y \in \mathcal{S}} \left| \frac{\partial^2 f_{u,l}}{\partial u^2}(y) \right| < \infty$. 
\end{assumption}

\begin{assumption}\label{assump::4}
$\{y_t\}_{t\in\mathbb{Z}}$ is $\alpha$-mixing. There exist some constant $C_1,C_2>0$ and $\beta>5/2$ such that $|\alpha(\ell)|\leq C_1 \ell^{-\beta}$ and $T^{(\beta+3/2)c-\beta/2+5/4}h^{-\beta/2-5/4}\to 0$ as $T\to\infty$. Moreover, $\left\{R_t^{(2)}\right\}_{t=1}^T$ is assumed to be strictly stationary.
\end{assumption}


Assumption~\ref{assump::1} is a common assumption used to ensure that each of the population Fr\'{e}chet means $f(t/T)$ in~\eqref{formula::trend component}, $\mu_Y$ in~\eqref{formula:: overall Frechet mean}, $g(t)$ in (\ref{formula::periodic component}), $\mu_{R,1}$ defined below (\ref{map2::remove periodic term}), and $\mu_{R,2}$ defined below (\ref{model::spheircal AR model}) exists and is unique. The first part of Assumption~\ref{assump::1} that requires $\sup_{\nu_1,\nu_2\in{\rm supp}(Y)}d_\mathcal{S}(\nu_1,\nu_2)\leq \pi/2$ is identical to Assumption (A1) in \cite{dai2022statistical}. As noted in \cite{dai2022statistical}, the support can be relaxed to $\sup_{\nu_1,\nu_2\in{\rm supp}(Y)}d_\mathcal{S}(\nu_1,\nu_2)\leq \pi$ if the dimension of the sphere is finite. If, for every $t=1,2,\ldots,T$, the distribution of $y_t$ is symmetric, like the von Mises--Fisher distribution (\citealt{mardia2009directional}), then the second part of Assumption~\ref{assump::1} holds when the first part of Assumption~\ref{assump::1} is satisfied. Assumption~\ref{assump::1} is crucial for the identification between the trend component $f$ and the periodic component $g$, see the discussion in Section~S3 in the supplementary material. Moreover, note that for compositional time series, the support after the square-root transformation is the hemisphere, and thus the first part of Assumption~\ref{assump::1} holds. However, for spherical time series, the first part of Assumption~\ref{assump::1} may be restrictive in applications featuring large trends that span a massive portion of the sphere. Relaxing this condition for the uniqueness of the Fr\'{e}chet mean remains an important direction for future work. Assumption~\ref{assump::2} is used to apply the results of \cite{silverman1978weak} and \cite{mack1982weak}, while Assumption~\ref{assump::3} is a standard distributional assumption for the smooth trend. Assumptions~\ref{assump::2} and \ref{assump::3} guarantee the asymptotic uniform equicontinuity of the cost function for the local Fr\'{e}chet regression used in Section~\ref{subsect::trend estimation}, and are consistent with the conditions (K0) and (R0) used in \cite{chen2022uniform} under the fixed design setting. Assumption~\ref{assump::4} is a condition on the $\alpha$-mixing coefficients of the spherical embedded time series, indicating that the coefficients exhibit polynomial decay. Assumption~\ref{assump::4} is a common condition used in nonlinear time series, see Chapter 6 in \cite{fan2003nonlinear}. As discussed at the beginning of this section, when the spherical AR model~\eqref{model::spheircal AR model} has a unique stationary solution, then Assumption~\ref{assump::4} naturally holds.

\subsection{Trend and periodic component estimation}\label{subsect::trend and periodic component est theory}

In this section, we derive the results of the uniform consistency rate for the trend and periodic component estimators in the STPD model~\eqref{final model}. Note that in this section, the final residual time series $\left\{R_t^{(2)}\right\}_{t=1}^T$ does not need to follow the spherical AR model in (\ref{model::spheircal AR model}), but it should satisfy the $\alpha$-mixing condition in Assumption~\ref{assump::4}, as discussed in Section~\ref{subsect::assumption}.

Note that the population counterpart of $\widehat{f}$ given in (\ref{formula::trend estimator}) is
\begin{align}\label{formula::population local Frechte}
    \widetilde{f}(u)=\argmin_{\nu\in\mathcal{S}}\widetilde{F}(\nu,u),~~ \widetilde{F}(\nu,u)=T^{-1}\sum_{t=1}^T \widehat{w}(t/T,u) Q(\nu,u_t,l_t),
\end{align}
where the weights $\widehat{w}(t/T,u)$ are given below (\ref{formula::trend estimator}) for $u\in[0,1]$ and $t\in\{1,2,\ldots,T\}$. As discussed in Section~\ref{subsect::trend estimation}, $\widetilde{f}(u)$ will not contain the information of the periodic pattern, supported by the following theorem.
\begin{theorem}\label{thm::oracle trend uniform rate}
    Under Assumptions~\ref{assump::1}--\ref{assump::4} in Section~\ref{subsect::assumption}, if the bandwidth $h\to0$ and $Th\to\infty$, then $\sup_{u\in[0,1]}d_\mathcal{S}\left(\widetilde{f}(u),f(u)\right)=O_p\left(h^2+(Th)^{-1}\right)$.
\end{theorem}
Theorem~\ref{thm::oracle trend uniform rate} indicates the population version of the local Fr\'{e}chet estimator, $\widetilde{f}$ in (\ref{formula::population local Frechte}), smooths out the periodic component, and will ensure that the estimated trend given by $\widehat{f}$ in (\ref{formula::trend estimator}) is asymptotically devoid of seasonal information. Theorem~\ref{thm::oracle trend uniform rate} also contains two sources of bias introduced by the local Fr\'{e}chet regression structure where $O(h^2)$ represents the classical smoothing bias for the local regression and $O((Th)^{-1})$ captures the seasonal filtering bias. Notably, the seasonal filtering bias aligns with the results in the real-valued time series setting, see Section~2.1 in the supplementary material of \cite{vogt2014nonparametric}.

Based on Theorem~\ref{thm::oracle trend uniform rate}, we establish the uniform consistency of the trend estimator $\widehat{f}(u)$ for $u \in [0, 1]$, as defined in (\ref{formula::trend estimator}). The following theorem provides the uniform convergence rate across the entire unit interval.

\begin{theorem}\label{thm::trend est}
Under Assumptions~\ref{assump::1}--\ref{assump::4} in Section~\ref{subsect::assumption}, if the bandwidth $h\to0$ and $Th^2/\log(1/h)\to\infty$, then $\sup_{u\in[0,1]}d_\mathcal{S}\left(\widehat{f}(u),f(u)\right)=O_p\left(h^2+\{Th/\log(1/h)\}^{-1/2}\right)$.
\end{theorem}

The uniform convergence rate presented in Theorem~\ref{thm::trend est} provides a theoretical guarantee that $\widehat{f}(u)$ is a consistent estimator of the true trend $f(u)$ across the entire domain. Theorem~\ref{thm::trend est} gives the same rate as for the local polynomial regression estimator under the $\alpha$-mixing condition in Euclidean space, see Theorem~6.5 in \cite{fan2003nonlinear}. The rate consists of two distinct components that reflect the classic bias-variance trade-off in nonparametric estimation. The first term $h^2$ represents the smoothing bias and arises from the local approximation of the function $f(u)$. The term $\{Th/\log(1/h)\}^{-1/2}$ accounts for the variance of the estimator where the presence of the factor $\log(1/h)$ is due to uniform convergence. Notably, under the conditions $h\to0$ and $Th^2/\log(1/h)\to\infty$, the bias term $O((Th)^{-1})$ in Theorem~\ref{thm::oracle trend uniform rate} is dominated by $O_p\left(\{Th/\log(1/h)\}^{-1/2}\right)$. Importantly, the uniform consistency of $d_\mathcal{S}\left(\widehat{f}(u),f(u)\right)$ in Theorem~\ref{thm::trend est} can be used in establish the identification between the trend component $f$ and the periodic component $g$, as discussed in Section~S3 in the supplementary material.

The asymptotic property of the period estimator $\widehat{\vartheta}_{\lambda_T}$, defined in (\ref{formula::period estimator}), is established in Theorem~\ref{thm::period estimation consistency}. For notational convenience regarding the asymptotic boundary conditions on the candidate search space and the regularisation sequence, we use $a_T\ll b_T$, equivalently $b_T\gg a_T$, to denote $a_T=o(b_T)$ for any arbitrary sequences $\{a_T\}$ and $\{b_T\}$.

\begin{theorem}\label{thm::period estimation consistency}
Suppose $\Theta_T \ll \min \left\{ Th^4, \frac{\log(1/h)}{h} \right\}$, the bandwidth $h\to0$, and $Th^2/\log(1/h)\to\infty$. Select the tuning parameter $\lambda_T$ such that $Th^2+T^{1/2}\{h/\log(1/h)\}^{-1/2}\ll \lambda_T \ll T$. Then $\widehat{\vartheta}_{\lambda_T}=\vartheta_0+o_p(1)$ as $T\to\infty$ under Assumptions~\ref{assump::1}--\ref{assump::4} in Section~\ref{subsect::assumption}.
\end{theorem}

Recall that the period estimation is conducted on the spherical time series $\left\{\widehat{R}_t^{(1)}\right\}_{t=1}^T$ where $\widehat{R}_t^{(1)} = \mathcal{M}_{\widehat{f}(t/T)\to y_t}\left(\widehat{\mu}_Y\right)$. Thus, the uniform convergence rate of the trend estimator $\widehat{f}$ plays a crucial role in the conditions for the upper bound $\Theta_T$ and the tuning parameter $\lambda_T$ in Theorem~\ref{thm::period estimation consistency}. Notably, $\Theta_T$ is allowed to diverge with a slower rate compared to the sample size $T$. Theorem~\ref{thm::period estimation consistency} establishes a broad range for the allowable growth rates of the tuning parameter $\lambda_T$. While any $\lambda_T$ locating within these theoretical bounds guarantees the large-sample consistency of $\widehat{\vartheta}_{\lambda_T}$, the finite-sample performance of $\widehat{\vartheta}_{\lambda_T}$ is sensitive to a chosen $\lambda_T$. This is also a motivation to propose the information criterion in (\ref{eq::BIC}) with theoretical support.

To develop the theoretical properties of the proposed information criterion, we decompose the tuning parameter space of $\lambda$ into three disjoint subsets as $\Lambda_0=\{\lambda:\widehat{\vartheta}_\lambda=\vartheta_0\}$, $\Lambda_+=\{\lambda:\widehat{\vartheta}_\lambda\in\mathcal{M}_{\vartheta_0},\widehat{\vartheta}_\lambda\neq\vartheta_0\}$, and $\Lambda_-=\{\lambda:\widehat{\vartheta}_\lambda\notin\mathcal{M}_{\vartheta_0}\}$. Here, $\mathcal{M}_{\vartheta_0}=\{\vartheta: \vartheta = k\vartheta_0,k=1,2,\ldots,1\leq \vartheta \leq \Theta_T\}$. These partitions, $\Lambda_0$, $\Lambda_+$, and $\Lambda_-$, directly map to scenarios where the global Fr\'{e}chet regression estimator $\widehat{g}_t(\vartheta,T)$ achieves the correct fit, suffers from overfitting, or is misspecified, respectively.

\begin{theorem}\label{thm::IC}
uppose $\Theta_T \ll \min \left\{ Th^4, \frac{\log(1/h)}{h} \right\}$, the bandwidth $h\to0$, and $Th^2/\log(1/h)\to\infty$. Then under Assumptions~\ref{assump::1}--\ref{assump::4} in Section~\ref{subsect::assumption}, $\mathbb{P}\left( \min_{\lambda\in\Lambda_+ \cup \Lambda_-} {\rm IC}_\lambda > {\rm IC}_{\lambda_T} \right) \to 1$ as $T\to\infty$ if (i) $\ell(T)\gg  h^2+\{Th/\log(1/h)\}^{-1/2}$ and (ii) $\ell(T)=o(1)$ as $T\to\infty$ for any reference tuning parameter $\lambda_T$ satisfying $Th^2+T^{1/2}\{h/\log(1/h)\}^{-1/2}\ll \lambda_T \ll T$. 
\end{theorem}

By Theorem~\ref{thm::IC}, any tuning parameter $\lambda$ that fails to correctly identify the true period $\vartheta_0$ is asymptotically excluded from being the optimal choice. This conclusion follows from the consistency results in Theorem~\ref{thm::period estimation consistency}, as for any sequence $\lambda_T$ satisfying $T^{1/2}\{h/\log(1/h)\}^{-1/2}\ll \lambda_T \ll T$, the estimator $\widehat{\vartheta}_{\lambda_T}$ converges in probability to the true parameter $\vartheta_0$ as $T \to \infty$. Consequently, Theorem~\ref{thm::IC} implies that if $\lambda$ belongs to the sets $\Lambda_+$ or $\Lambda_-$, the resulting information criterion ${\rm IC}_\lambda$ will exceed the reference level ${\rm IC}_{\lambda_T}$ with a probability approaching one. Given that the optimal selector is defined as $\widehat{\lambda} = \argmin_\lambda {\rm IC}_\lambda$, the inequality ${\rm IC}_{\widehat{\lambda}} \leq {\rm IC}_{\lambda_T}$ must hold. This ensures that $\widehat{\lambda}$ is contained within $\Lambda_0=\{\lambda:\widehat{\vartheta}_\lambda=\vartheta_0\}$. Theorem \ref{thm::IC} leads to a specific consideration of the information criterion as 
\[
{\rm IC}_\lambda = \log \{ {\rm RSS}(\widehat{\vartheta}_\lambda) / T \} + \widehat{\vartheta}_\lambda \max \left\{ h^2 \log(1/h), \sqrt{\frac{\log^2(1/h)}{Th}} \right\}.
\]

Combining Theorems~\ref{thm::period estimation consistency} and \ref{thm::IC} establishes the consistency of $\widehat{\vartheta}_{\widehat{\lambda}}$ with $\widehat{\lambda}$ selected via (\ref{formula::lambda selection}), shown in Corollary \ref{cor::final period est} below.

\begin{corollary}\label{cor::final period est}
Suppose that the conditions in Theorem~\ref{thm::IC} hold.  Then $\widehat{\vartheta}_{\widehat{\lambda}}-\vartheta_0=o_p(1)$ with $\widehat{\lambda}$ given by (\ref{formula::lambda selection}).
\end{corollary}

We now derive the asymptotics of the periodic component estimator $\widehat{g}$ given by (\ref{formula::periodic component estimator}).

\begin{theorem}\label{thm::periodic component est consistency}
Suppose that the conditions in Theorem~\ref{thm::IC} hold, then
\[
\max_{t=1,2,\ldots,T} d_\mathcal{S}(\widehat{g}(t),g(t)) = O_p\left( h^2 + \{Th/\log(1/h)\}^{-1/2} \right).
\]
\end{theorem}

The estimator $\widehat{g}$ is obtained via the spherical time series $\left\{\widehat{R}_t^{(1)}\right\}_{t=1}^T$ where $\widehat{R}_t^{(1)} = \mathcal{M}_{\widehat{f}(t/T)\to y_t}\left(\widehat{\mu}_Y\right)$, meaning that $\widehat{g}$ is a statistic with estimated parameters involving $\widehat{f}$. Consider an oracle estimator $\widetilde{g}(t)=\widetilde{g}_t\left(\widehat{\vartheta}_{\widehat{\lambda}},T\right)$ where $ \widetilde{g}_t(\vartheta, T) = \argmin_{\nu\in\mathcal{S}} T^{-1}\sum_{i=1}^T s^{(t)}_i d_\mathcal{S}^2\left(\nu,R_i^{(1)}\right)$. We show that $\max_{t=1,2,\ldots,T} d_\mathcal{S}(\widehat{g}(t),\widetilde{g}(t)) = O_p\left( h^2+ \{Th/\log(1/h)\}^{-1/2} \right)$ and $\max_{t=1,2,\ldots,T} d_\mathcal{S}(\widetilde{g}(t),g(t)) = O_p\left( T^{-1/2} \right)$. Not surprisingly, the uniform convergence rate of the local regression estimator $\widehat{f}$ is slower than that of the global regression estimator $\widetilde{g}$. Thus, $\max_{t=1,2,\ldots,T} d_\mathcal{S}(\widehat{g}(t),\widetilde{g}(t))$ is the leading term, leading to the uniform convergence rate in Theorem~\ref{thm::periodic component est consistency}.

\subsection{Spherical AR model parameter estimation}\label{subsect::AR coef est theory}

Here, we present the convergence rate of the spherical AR coefficient estimators.

\begin{theorem}\label{thm::AR parameter est consist}
Suppose that the conditions in Theorem~\ref{thm::IC} hold, let 
\[
\Lambda_\varrho = \begin{pmatrix}
\varrho_0 & \varrho_1 & \cdots &\varrho_{p-1} \\
\varrho_1 & \varrho_0 & \cdots &\varrho_{p-2} \\
\vdots & \vdots && \vdots \\
\varrho_{p-1} & \varrho_{p-2} & \cdots &\varrho_0 \\
\end{pmatrix}.
\]
If ${\rm det}(\Lambda_\varrho)\neq 0$, then $\|\widehat{\varphi}-\varphi\|_E=O_p( h^2+ \{Th/\log(1/h)\}^{-1/2} )$ where $\widehat{\varphi}=(\widehat{\varphi}_1,\widehat{\varphi}_2,\ldots,\widehat{\varphi}_p)^\top$, and $\varphi=(\varphi_1,\varphi_2,\ldots,\varphi_p)^\top$.
\end{theorem}
The estimation of the spherical AR coefficient is calculated via~\eqref{formula::AR coef estimation}, which involves all estimators $\widehat{\vartheta}_{\widehat{\lambda}},\widehat{g},\widehat{f}$, and $\widehat{\mu}_Y$, and thus gives the slower convergence rate in Theorem~\ref{thm::AR parameter est consist}, compared to the rate $O_p(T^{-1/2})$ in \citet[][Corollary~1]{zhu2024spherical}.

\section{Simulation}\label{sect::simulation}

This section first introduces the data generating process briefly with details given in Section~S7.1 in the supplementary material. Here, we present the simulation results on prediction in Section~\ref{subsect::simulation prediction}, and the simulation results on estimation can be found in Section~S7.2 in the supplementary material. For the ease of reproducibility, computational code is available at \url{https://github.com/d5rhtkp4kc-alt/STPD}. The implementations for the global and local Fr\'{e}chet regression are largely built upon the functions provided in the \Rlogo\ package \texttt{frechet} \citep{frechet2023}.

Data generation consists of three steps where we first generate stationary spherical time series following the spherical AR structure in (\ref{model::spheircal AR model}), then introduce a periodic component into the stationary spherical time series, and finally add a trend component. In the following, we mainly focus on the space $\mathcal{S}^6$ where a much higher dimension case can be found in Section~S7 of the supplementary material.

\begin{figure}[!htb]
\centering
\includegraphics[width=0.85\linewidth]{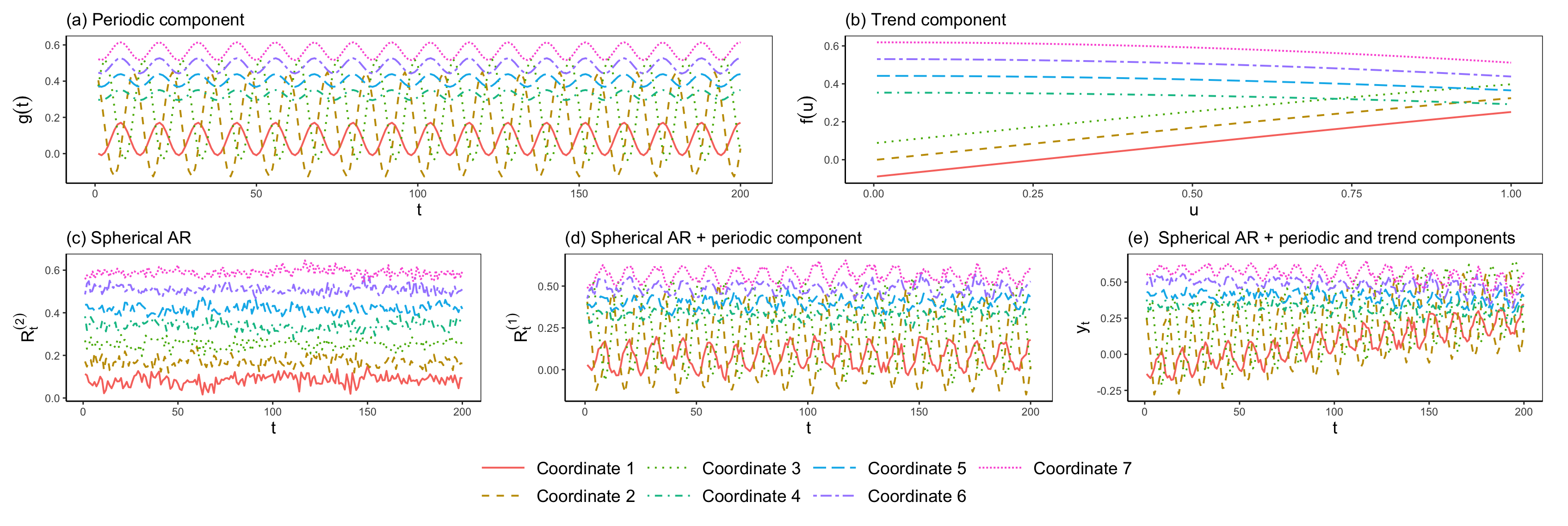}
\caption{The timeline of (a) the generated periodic component $g(t)$, (b) the generated trend component $f(u)$, (c) the simulated stationary spherical residuals $R_t^{(2)}$ using a AR coefficient vector $(0.3,-0.1,0.4)^\top$, (d) the simulated spherical residuals $R_t^{(1)}$ by integrating the generated periodic component $g(t)$ to $R_t^{(2)}$, and (e) the simulated spherical time series $y_t$ by integrating the generated trend component $f(u)$ to $R_t^{(1)}$, where $t=1,2,\ldots,200$ and $u=t/200$. Each line represents a coordinate of the simulated spherical vector.}\label{fig::simulated components}
\end{figure}

Figure~\ref{fig::simulated components} visualizes the generated spherical vectors, $g(t), f(u),R_t^{(2)},R_t^{(1)},y_t$, in a typical simulation replicate. The simulation parameters are set to a sample size $T=200$, AR order $p=3$ with coefficients $\psi_1=0.3, \psi_2=-0.1, \psi_3=0.4$, and a period $\vartheta_0=12$. Figure~\ref{fig::simulated components}(a) shows that the generated periodic component $g(t)$ has a period of 12 as expected, while the trend component $f(u)$ in Figure~\ref{fig::simulated components}(b) displays increasing trends for the first three coordinates of the spherical vector while the remaining coordinates exhibit decreasing trends due to the constraint caused by the spherical structure. Figure~\ref{fig::simulated components}(c) indicates that the simulated spherical residuals $R_t^{(2)}$ are stationary, while after imposing the periodic component $g(t)$ into $R_t^{(2)}$, a clear periodic pattern of $R_t^{(1)}$ can be observed in Figure~\ref{fig::simulated components}(d). Finally, after integrating the trend component $f(t/T)$ into $R_t^{(1)}$, Figure~\ref{fig::simulated components}(e) confirms that the spherical time series $y_t$ shows a periodic pattern and a trend behavior. In the following, we set the true period $\vartheta_0=12$.

\subsection{Simulation results on prediction}\label{subsect::simulation prediction}

This section evaluates the finite-sample prediction performance of the proposed TPSAR model in Section~\ref{sect::spherrical AR setup}, compared to four baselines covering the Spherical AR (SAR), Differencing-based Spherical AR (DSAR) models \citep{zhu2024spherical}, two LSTM models covering the standard LSTM model (LSTM-S) and the periodically-adjusted LSTM model (LSTM-P), and the TimePFN model. For a spherical time series $\left\{y_t\right\}_{t=1}^T$, the SAR model is to apply the model (\ref{model::spheircal AR model}) directly to the original time series by letting $\Xi_t=y_t\ominus\mu_Y$ for $t\in\{1,2,\ldots,T\}$, while the DSAR model is the one considering $\Xi_t=y_t\ominus y_{t-1}$ for $t\in\{2,3,\ldots,T\}$. Additional information of the LSTM models and the TimePFN model used in numerical studies is provided in Section~S6 in the supplementary material.

Here, the replica number is set to 200 and the prediction performance is evaluated in the following way. For each replicate $k\in\{1,2,\ldots,200\}$ and a given spherical AR order $p\in\{1,2,3\}$, we generate spherical time series $\left\{y_t^{(k)}\right\}_{t=1}^T$ following the same data generation procedure in Section~S7.2 whose details are given in Section~S7.1 in the supplementary material. For a given $\varkappa\in(0,1)$, we use $\left\{y_t^{(k)}\right\}_{t=1}^{\lfloor T\varkappa\rfloor}$ as a training sample and $\left\{y_t^{(k)}\right\}_{t=\lfloor T\varkappa\rfloor+1}^T$ as a test sample. Then, for each model and each forecast horizon $m \in \{T-\lfloor T\varkappa\rfloor,T-\lfloor T\varkappa\rfloor-1, \ldots, 1\}$, we define a rolling training window $\left\{y_t^{(k)}\right\}_{t=T-\lfloor T\varkappa\rfloor-m+1}^{T-m}$ to fit the model, with the AR order selected up to a maximum of $20$ using the rolling-window selection criterion in the Appendix. We then generate $m$-step-ahead forecasts following the intrinsic prediction scheme detailed in Section~\ref{sect::prediction}, denoted as $\left\{\widehat{y}_t^{(k)}\right\}_{t=T-m+1}^T$. The forecast accuracy at each replicate $k$ is measured by the average geodesic distance on the sphere as ${\rm Dist}(k,m)=m^{-1}\sum_{t=T-m+1}^T d_\mathcal{S}\left(\widehat{y}_t^{(k)},y_t^{(k)}\right)$ for every forecast horizon $m$. Finally, we report the average prediction error for every $m$ as $(200)^{-1}\sum_{k=1}^{200}{\rm Dist}(k,m)$.
\begin{figure}[!htb]
\centering
\includegraphics[width=0.85\linewidth]{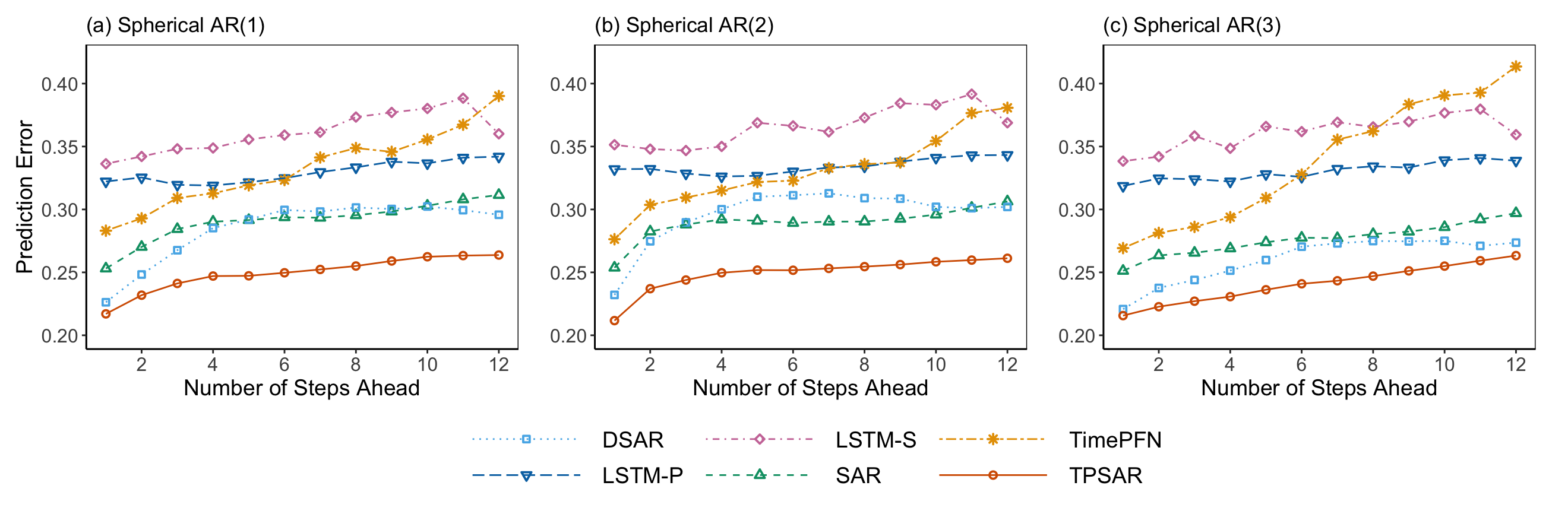}
\caption{The average prediction error for $m$-step-ahead predictions calculated via the average angle (radian) between the simulated true observations and the predictions.  The observations are located within $\mathcal{S}^6$. The number of steps ahead $m$ ranges from~1 to 12, the sample size being 120, and the simulated final residual process is generated via a spherical AR model with (a) order being~1, (b) order being~2, and (c) order being~3. Six models are evaluated: the spherical AR (SAR) model, the differencing-based spherical AR model (DSAR), the standard LSTM (LSTM-S) model, the periodically-adjusted LSTM (LSTM-P) model, the TimePFN model, and the proposed TPSAR model.}\label{fig::simulation predict T=120}
\end{figure}

Figure~\ref{fig::simulation predict T=120} displays the average prediction errors for the proposed TPSAR model alongside the SAR, DSAR, LSTM-S, LSTM-P, and TimePFN models when the sample size $T=120$, across three distinct generating processes for the spherical residual series. As anticipated, the prediction error exhibits a monotonic increase as the forecast horizon $m$ extends for the TPSAR model. The results also indicate that the TPSAR model consistently outperforms all the five baselines across all scenarios and forecast horizons tested. Given that the simulated spherical time series are influenced by deterministic trend and periodic components, the resulting non-stationarity violates the fundamental assumptions of the SAR model. Furthermore, while the DSAR model attempts to mitigate non-stationarity through differencing, it appears insufficient for decoupling complex deterministic signals from the underlying stochastic process. Notably, the DSAR model occasionally exhibits higher error rates than the SAR model. For short-term horizons, e.g., $m=1$, the performance gap between the three AR methodologies narrows. This is likely due to the setting that the maximum AR candidate order is $20$, which enables the SAR and DSAR models to exploit local information through high-order lags. However, as the forecast horizon $m$ increases, the predictive advantage of the TPSAR model becomes more pronounced. In contrast to the performance of the AR models, the LSTM and TimePFN models exhibit a significant degradation in predictive accuracy, mainly due to the small sample size.

Similar patterns can be observed for a larger sample size $T=300,600$ in Figure~S3 in the supplementary material. Notably, Figure~S3 in the supplementary material also shows that when the sample size increases, the average prediction error for the proposed TPSAR model decreases. Furthermore, with increasing sample sizes, the LSTM and TimePFN models achieve better prediction performance and surpass both the SAR and DSAR models when $T=600$, yet the proposed TPSAR model continues to yield superior performance in all scenarios.

\section{Empirical data application}\label{sect::real data}

\subsection{U.S. electricity generation compositions}\label{subsect::real data--composition}

We investigate the monthly structural dynamics of U.S. electricity production, sourcing raw data via \url{https://www.eia.gov/electricity/data/browser/}. We transform these raw data into a sequence of compositional vectors, where individual components quantify the relative fractional share of distinct fuel types to the aggregate net generation. Adopting the data preprocessing strategy  in \cite{xu2025quantifying}, we group granular fuel classifications into seven macro-categories such as coal, petroleum (including petroleum liquids and petroleum coke), gas (natural and other gases), nuclear, conventional hydroelectric, renewables (wind, geothermal, and biomass), and solar (encompassing small-scale solar photovoltaic and all utility-scale solar). This preprocessing gives the result that the space of the compositions is a six-simplex $\Delta^6=\{(\delta_1,\ldots,\delta_7)^\top\in\mathbb{R}^7:\sum_{l=1}^7 \delta_l = 1~{\rm and}~\delta_p>0~{\rm for}~p=1,\ldots,7\}$. Our data set consists of $T=276$ monthly observations spanning from January 2001 to December 2023. Following the discussion in Section~\ref{subsect::spherical embedded ts}, we apply the pointwise square root transformation for compositional time series, resulting in a spherical time series. We first apply the STPD model (\ref{final model}) to decompose the spherical time series into a trend component, a periodic component and a final residual time series, with an estimation procedure detailed in Section~\ref{sect::component est procedure}. Here, the upper bound of the period candidate used in the periodic component estimation is set to $\Theta_T=40$.
\begin{figure}[!htb]
\centering
\includegraphics[width=0.85\linewidth]{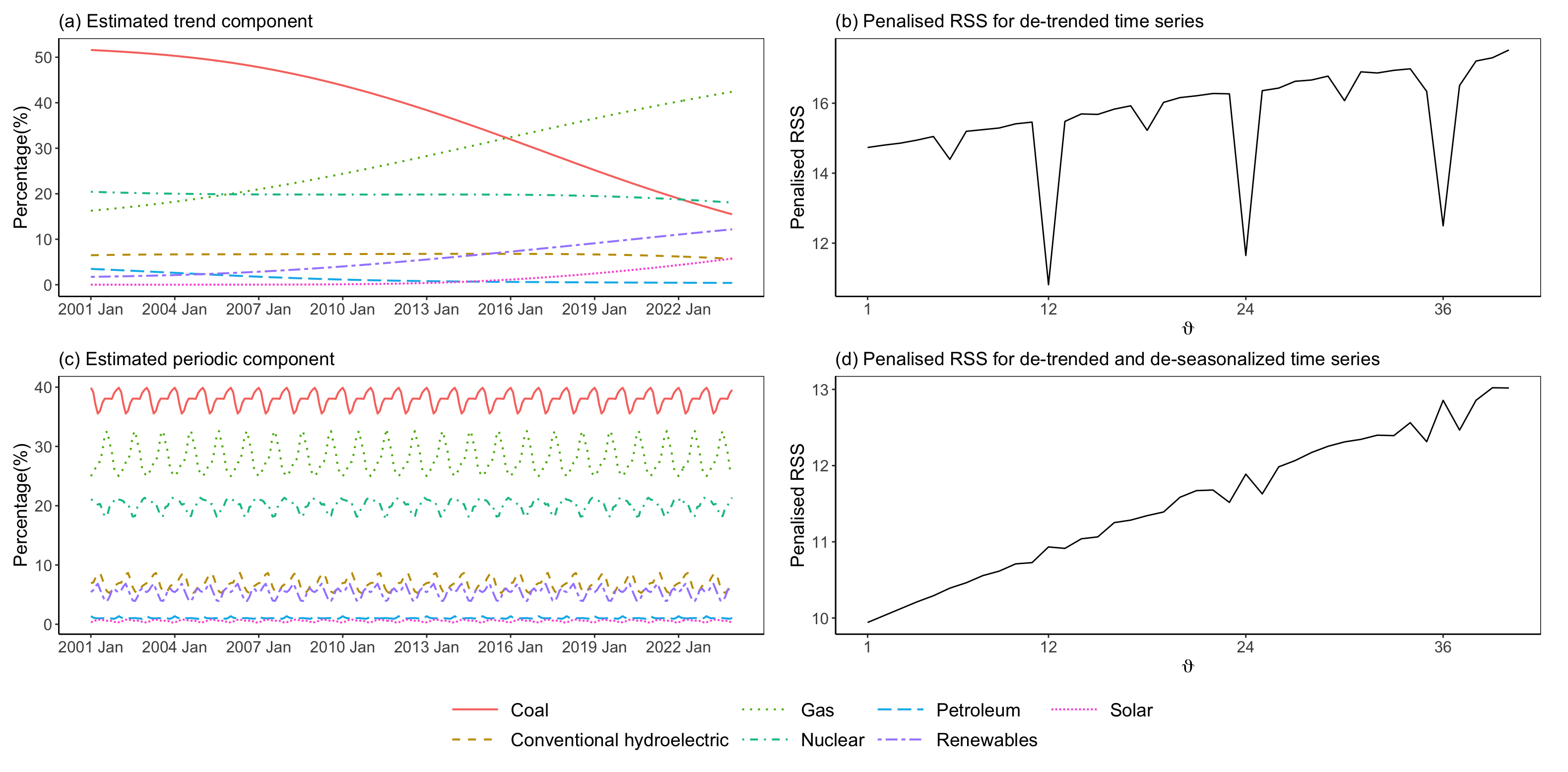}
\caption{(a) The timeline of the estimated trend component of the monthly electricity generation compositional time series in U.S. from January 2011 to December 2023, (b) the plot of the penalized RSS with the selected tuning parameter for the de-trended compositional time series, (c) the timeline of the estimated periodic component of the compositions, and (d) the plot of the penalized RSS with the selected tuning parameter for the de-trended and de-seasonalized compositional time series. Each line in (a) and (c) represents the percentage of the corresponding energy resource categories covering Coal, Petroleum, Gas, Nuclear, Conventional hydroelectric, Renewables, and Solar.}\label{fig::US energy estimation}
\end{figure}

Figure~\ref{fig::US energy estimation} summarizes the results of the trend and periodic component estimation procedure, and we apply the square transformation for the estimated components, ensuring for every time point that their elements are located within the six-simplex $\Delta^6$. Figure~\ref{fig::US energy estimation}(a) illustrates the evolution of the estimated trend component, where each line represents the percentage of the corresponding categories of energy resources. A significant decline in percentage of coal generation can be observed in Figure~\ref{fig::US energy estimation}(a), contrasted by a steady increase in gas, solar, and other renewables. Figure~\ref{fig::US energy estimation}(b) displays the penalized RSS for the original time series after trend removal, and the global minimum clearly identifies an estimated period of 12. Given the monthly frequency of the data, this suggests a prominent intra-year cycle. Following the estimated period, the estimated periodic component is shown in Figure~\ref{fig::US energy estimation}(c). In alignment with \cite{xu2025quantifying}, the share of natural gas peaks between July and September, reflecting its dominant role in meeting the peak summer electricity demand. Conversely, the proportion of nuclear power decreases during these months. This dip is largely attributable to elevated summer water temperatures, which reduce cooling efficiency of nuclear power plants and necessitate lower power outputs to maintain safe operational limits. To validate the extraction of the estimated periodic component, Figure~\ref{fig::US energy estimation}(d) shows the penalized RSS for the final de-trended and de-seasonalized residuals. The global minimum in this plot indicates that the period is 1 and thus the periodic signal has been successfully removed. It is worth noting that although total energy generation exhibits an annual periodic pattern, peaking in summer and winter due to elevated demand, the composition of energy sources does not necessarily follow this yearly cycle. Indeed, the relative proportions of each energy source can remain constant, even as the aggregate generation volume fluctuates periodically. This yearly cycle of the energy generation compositional time series has not been fully explored in the literature.

\begin{figure}[!htb]
\centering
\includegraphics[width=0.85\linewidth]{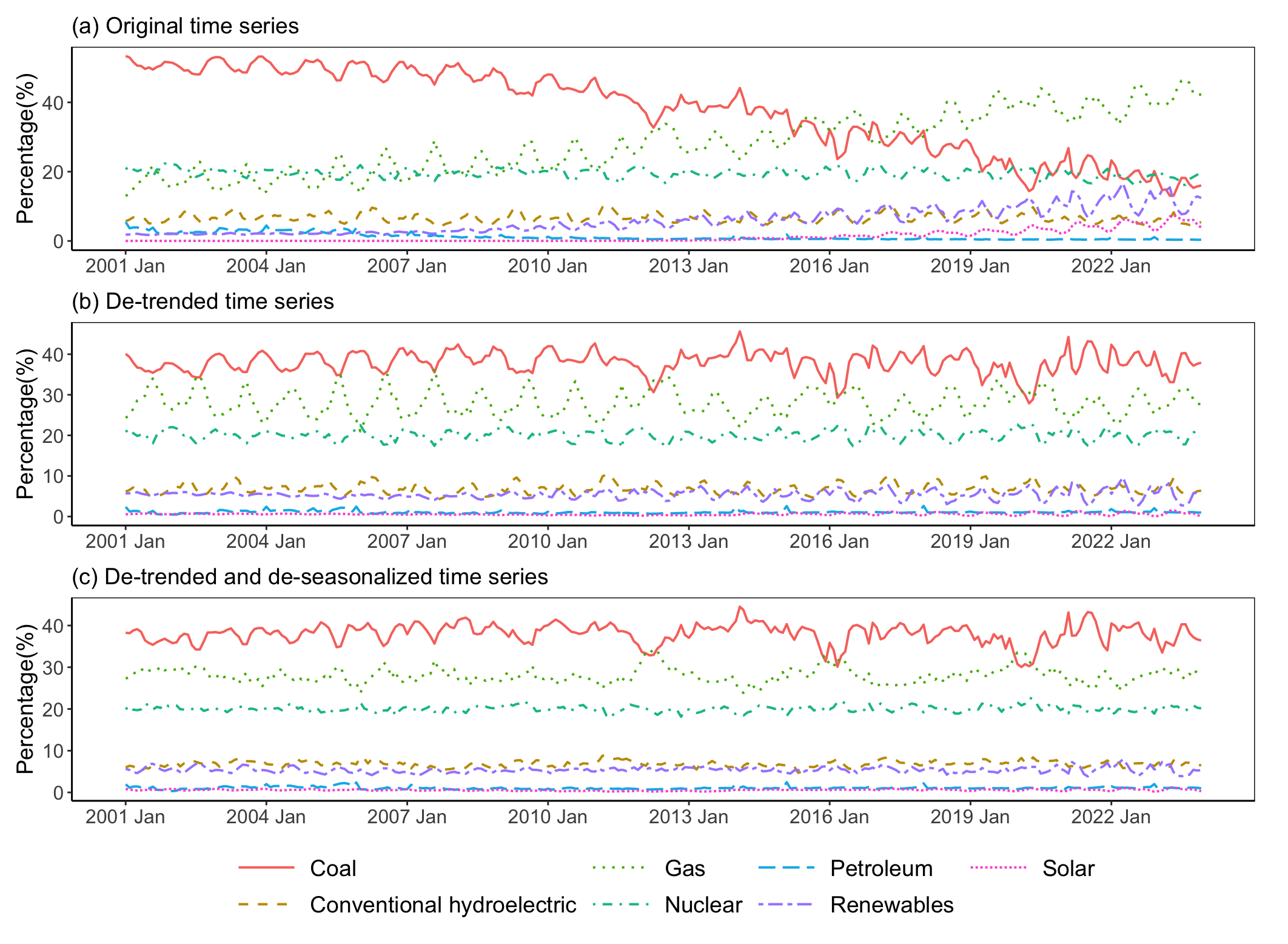}
\caption{The timeline of (a) U.S. electricity generation compositions, (b) the de-trended compositions, (c) the de-trended and de-seasonalized compositions, from January 2011 to December 2023. Each line represents the percentage of the corresponding energy resource categories covering Coal, Petroleum, Gas, Nuclear, Conventional hydroelectric, Renewables, and Solar.}\label{fig::US energy spherical data}
\end{figure}

Figure~\ref{fig::US energy spherical data} illustrates the effect of the proposed trend and periodic component removal applied in the U.S. electricity generation compositional time series. Figure~\ref{fig::US energy spherical data}(a) displays the raw compositional data, revealing a clear evolutionary trend that aligns with the estimated component in Figure~\ref{fig::US energy estimation}(a). While a periodic pattern is visible in the raw compositional time series, which is most notable in gas, it remains relatively subtle for other energy sources. The effectiveness of trend removal is demonstrated in Figure~\ref{fig::US energy spherical data}(b). For example, the pronounced decline in coal generation percentages seen in the original series is successfully neutralized, with the de-trended line stabilizing around 35\%. This removal effectively isolates the periodic structure, making seasonal fluctuations significantly more apparent than in the raw data. Finally, Figure~\ref{fig::US energy spherical data}(c) presents the residuals after removing both trend and periodic components. The removal of seasonality is evident, particularly for gas and petroleum, and is consistent with the penalized RSS shown in Figure~\ref{fig::US energy estimation}(d). The resulting residual series exhibits stationarity, especially in contrast to the original and de-trended series in Figures~\ref{fig::US energy spherical data}(a) and (b).

Finally, we assess the predictive performance of the proposed TPSAR model in Section~\ref{sect::spherrical AR setup} against the SAR model, the DSAR model, two LSTM models, and the TimePFN model, whose details are given in Section~S8.1 of the supplementary material and the proposed TPSAR model performs better in most scenarios.

\subsection{New York City Citi Bike trip volume shapes}\label{subsect::real data--distribution}

The New York City Citi Bike program provides historical trip records with second-level temporal resolution, accessible through \url{https://citibikenyc.com/system-data}. Our analysis investigates the daily profiles of trip volume, which we model as a distributional time series. Specifically, for each day, we aggregate the total number of trips at the midpoint of each 30-minute interval, yielding a daily functional representation of system-wide usage. To isolate the shape of the temporal dynamics of usage patterns in total daily commuting, we normalize these functions into probability density functions. We focus on $T=214$ daily observations that span from March 2019 to September 2019. Following the discussion in Section~\ref{subsect::spherical embedded ts}, we apply the functional square root transformation for distributional time series, resulting in a spherical time series. Utilizing the STPD model (\ref{final model}) and following the procedure in Section~\ref{sect::component est procedure} by setting the upper bound of the period candidate used in the periodic component estimation be $\Theta_T=25$ , we decompose the spherical time series into a trend component, a periodic component, and a final residual time series.

\begin{figure}[!htb]
\centering
\includegraphics[width=0.85\linewidth]{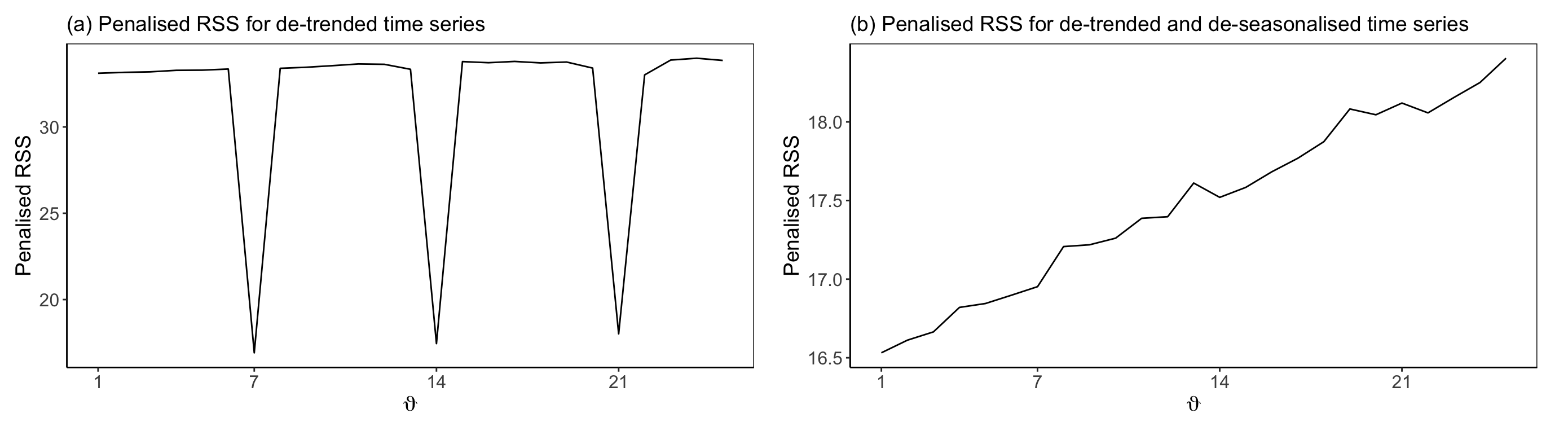}
\caption{The plots of the penalized RSS with the selected tuning parameter for (a) the de-trended distributional time series and (b) the de-trended and de-seasonalized distributional time series.}\label{fig::NYC penalized RSS}
\end{figure}

Figure~\ref{fig::NYC penalized RSS} illustrates the penalized RSS utilized for periodicity detection. Specifically, Figure~\ref{fig::NYC penalized RSS}(a) shows the penalized RSS calculated after trend removal, while Figure~\ref{fig::NYC penalized RSS}(b) depicts the penalized RSS following the extraction of both the trend and periodic components. As shown in Figure~\ref{fig::NYC penalized RSS}(a), the penalized RSS identifies the estimated period of 7 for the daily trip volume profiles. This result reveals a clear intra-week periodic pattern, consistent with the empirical findings in \cite{xu2025robust}. Furthermore, the penalized RSS in Figure~\ref{fig::NYC penalized RSS}(b) confirms the efficacy of the periodic component removal as the estimated period becomes one, indicating that there is no periodic pattern.

\begin{figure}[!htb]
\centering
\includegraphics[width=0.85\linewidth]{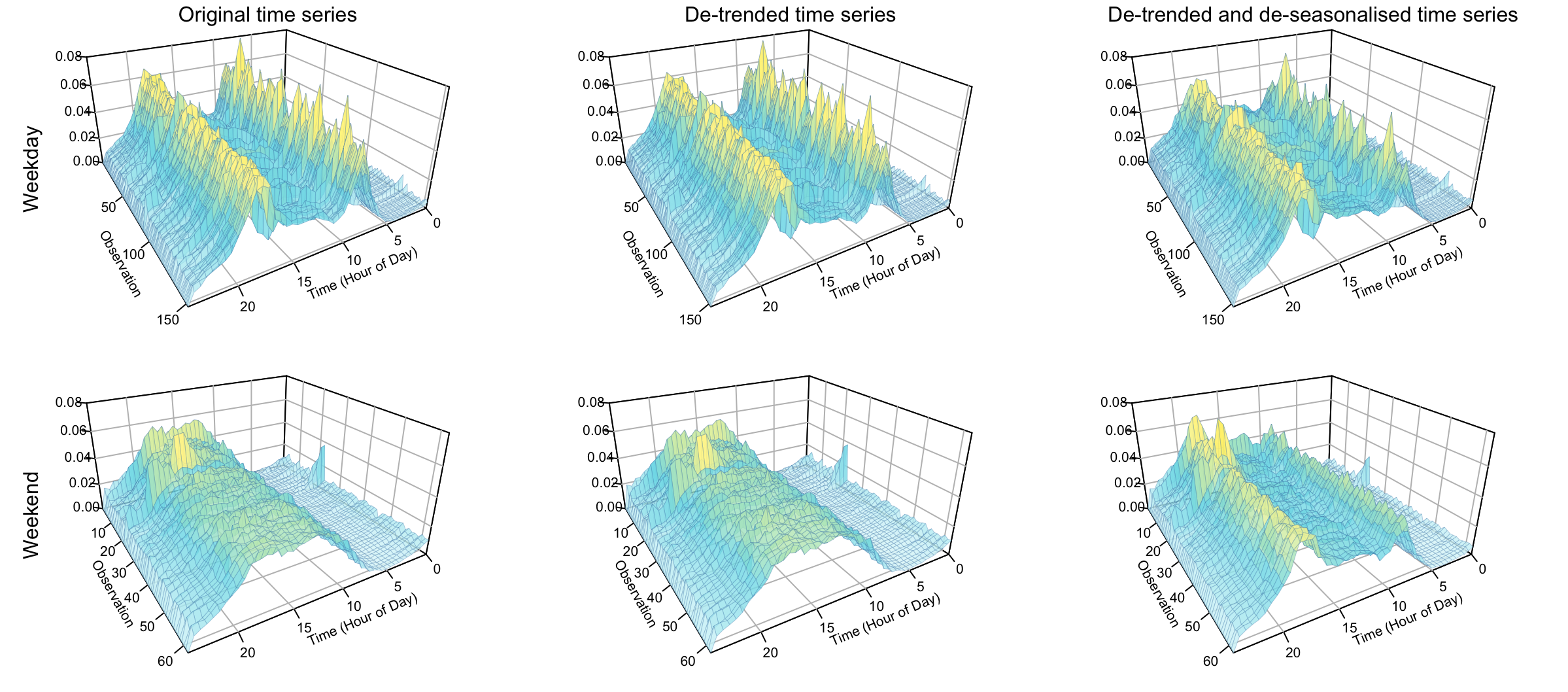}
\caption{ The top row displays data for weekdays, while the bottom row represents weekends in New York City City Bike sharing system from March 2019 to October 2019. The first column shows the original distributional time series, the second column presents the de-trended time series, and the last column illustrates the final de-trended and de-seasonalized time series.}\label{fig::NYC TPDM weekday weekend group}
\end{figure}

To further illustrate the evolution of the periodic structures throughout the decomposition procedure, we split the time series into two distinct segments corresponding to weekdays and weekends, while preserving their chronological ordering. The corresponding trajectories are visualized in Figure~\ref{fig::NYC TPDM weekday weekend group} while the visualization of the whole time series without partition can be found in Figure~S6 in the supplementary material. The first column of Figure~\ref{fig::NYC TPDM weekday weekend group} highlights a significant difference between the raw weekday and weekend trip volume profiles, where weekday trajectories are consistently bimodal, featuring localized peaks around 9am and 5pm, while weekend profiles are predominantly unimodal, exhibiting a sustained plateau between 10 am and 4 pm. As observed in the second column, this intra-week periodic pattern persists after the removal of the underlying trend. However, following the subsequent extraction of the periodic component, the third column demonstrates that the structural disparity between the weekday and weekend profiles is effectively neutralized, yielding a homogeneous residual process. Similar patterns can be observed in a further partition of the time series into individual days of the week provided in Figures~S7 and S8 of the supplementary material. Additional analysis of the estimated trend and periodic components can be found in Section~S8.2 in the supplementary material.

Finally, we evaluate the out-of-sample predictive performance of the proposed TPSAR model against the SAR model, the DSAR model, two LSTM models, and the TimePFN model, whose details are given in Section~S8.2 of the supplementary material, and the TPSAR model achieves superior accuracy in most scenarios. 

\section{Conclusion}\label{sec:conclu}

Deterministic trends and periodic patterns are fundamental characteristics of both Euclidean and non-Euclidean time series. Despite their prevalence, there remains a notable paucity of methodological frameworks capable of capturing such nonstationary structures within spherically embedded data. This paper addresses this lacuna by introducing a unified geometric framework for the analysis and forecasting of spherically embedded time series subject to unknown trend and periodic components. Our work fills a critical gap in the existing literature, which has predominantly focused on some certain stationary processes that fail to capture latent trends and periodic patterns. The cornerstone of our methodology is the STPD model, enabling systematic decomposition of complex spherically embedded time series into interpretable deterministic trend, periodic component, and stationary stochastic residuals. The effectiveness and versatility of the proposed framework are demonstrated through extensive simulation studies and real data applications involving compositional time series of electricity generation and volume profiles of cycle trips. In particular, our applications to these two real-world datasets demonstrate that this framework not only improves predictive accuracy, but also reveals latent structural dynamics with meaningful interpretation.

It is worth noting that the proposed spherical trend-periodicity decomposition model can be extended to general geodesic spaces. The geometric interpretation of the proposed removal operator relies on the geodesic transport map. Therefore, if a suitable transport operator can be defined for a given metric space, such as parallel transport on general Riemannian manifolds, an analogous trend-periodicity decomposition model can be constructed using the same underlying principles.

\section*{Supplementary Material}

The supplementary material includes a discussion on computational complexity of the proposed four-stage procedure, the failure when using several extrinsic ways to remove the trend and periodic components, a discussion on a possible tangent-space-based approach, the identification between the trend and periodic components, several technical lemmas, proofs of the main results, additional information of the LSTM and TimePFN models, the data generation process and the additional simulation results, and additional results for empirical data analysis. 

\if0\blind
{
\section*{Acknowledgement}

The authors thank the two reviewers and the Associate Editor for their insightful comments. We are also grateful for the feedback received from participants at various conferences, including but not limited to the International Conference on Contemporary Statistics and Data Science, the New Researchers Conference Asia, and the University of Sydney Data Science Symposium. This research was supported by an Australian Research Council Future Fellowship (FT240100338) and undertaken with assistance from computational resources provided by the Australian Government through the National Computational Infrastructure (NCI) under the Macquarie University Merit Allocation Scheme.
} \fi

\if1\blind
{

} \fi

\section*{Appendix}
Here, we present the rolling-window cross-validation criterion. Suppose that one has a spherical embedded time series $\{z_t\}_{t=1}^T$ and a target model with a hyperparameter, for instance, the local Fr\'{e}chet model $\widehat{f}$ in (\ref{formula::trend estimator}) with a candidate bandwidth $h$, or the spherical AR model in (\ref{model::spheircal AR model}) with a candidate order $p$. Given $\varkappa\in(0,1)$, for every $t_1=1,2,\ldots,\lfloor T\varkappa\rfloor$, we use $\{z_{\lfloor T\varkappa\rfloor-t_1+1},z_{\lfloor T\varkappa\rfloor-t_1+2},\ldots,z_{T-t_1}\}$ to train the target model with the hyperparameter candidate. Based on the fitted model and the training data, we calculate the prediction $\widehat{z}_{T-t_1+1}$ and the hyperparameter is selected so that the prediction error $(\lfloor T\varkappa\rfloor)^{-1}\sum_{t_1=1}^{\lfloor T\varkappa\rfloor}d_\mathcal{S}(\widehat{z}_{T-t_1+1},z_{T-t_1+1})$ is the minimum.


\begin{singlespace}
\bibliographystyle{apalike}
\bibliography{reference}
\end{singlespace}

\end{document}